\documentclass[a4paper,USenglish,cleveref, autoref, thm-restate,nolineno]{socg-lipics-v2021}
\graphicspath{ {images/} }

\newenvironment{proof-sketch}{\noindent{\bf The proof sketch.}\hspace*{1em}}{\qed\bigskip}

\newcommand{\sS}{\mathcal{S}} % for a set of elements
\newcommand{\sR}{\mathcal{R}} % for a set of ranges
\newcommand{\sP}{\mathcal{P}} % for a set of points
\newcommand{\rR}{\mathscr{R}} % for a range
\newcommand{\bR}{\mathbb{R}} % real field
\newcommand{\C}{\boldsymbol{C}} % hypercube
\newcommand{\U}{\boldsymbol{U}} % hyper unit cube
\newcommand{\bm}{\boldsymbol{m}} % max # params
\newcommand{\bb}{\boldsymbol{b}} % max # params in 2D
\newcommand{\bbb}{\mathcal{B}} % binom{\bb}{2}
\newcommand{\D}{\Delta} % Delta
\newcommand{\ba}{\mathbf{a}}
\newcommand{\bv}{\mathbf{v}}
\newcommand{\bw}{\mathbf{w}}
\newcommand{\bp}{\mathbf{p}}
\newcommand{\bc}{\mathbf{c}}
\newcommand{\bd}{\mathbf{d}}
\newcommand{\bg}{\boldsymbol\gamma}

\newcommand{\tinyo}{\scriptscriptstyle o}
\newcommand{\domega}{\overset{\tinyo}{\Omega}}
\newcommand{\dtheta}{\overset{\tinyo}{\Theta}}
\newcommand{\dO}{\overset{\tinyo}{O}}
\newcommand{\etal}{\textit{et al.}}

\newcommand{\ignore}[1]{}

\usepackage{subfiles}
\usepackage{physics}
\usepackage{setspace}

\title{On Semialgebraic Range Reporting} %TODO Please add

\author{Peyman Afshani}{Aarhus University, Aarhus, Denmark}{peyman@cs.au.dk}{}{}%TODO mandatory, please use full name; only 1 author per \author macro; first two parameters are mandatory, other parameters can be empty. Please provide at least the name of the affiliation and the country. The full address is optional. Use additional curly braces to indicate the correct name splitting when the last name consists of multiple name parts.

\author{Pingan Cheng}{Aarhus University, Aarhus, Denmark}{pingancheng@cs.au.dk}{}{}

\authorrunning{P. Afshani and P. Cheng} %TODO mandatory. First: Use abbreviated first/middle names. Second (only in severe cases): Use first author plus 'et al.'

\Copyright{P. Afshani and P. Cheng} %TODO mandatory, please use full first names. LIPIcs license is "CC-BY";  http://creativecommons.org/licenses/by/3.0/

 %TODO mandatory: Please choose ACM 2012 classifications from https://dl.acm.org/ccs/ccs_flat.cfm 

 %TODO mandatory; please add comma-separated list of keywords

\category{} %optional, e.g. invited paper

\relatedversion{} %optional, e.g. full version hosted on arXiv, HAL, or other respository/website
\funding{Supported by DFF (Det Frie Forskningsr\aa d) of Danish Council for Independent Research under grant ID DFF$-$7014$-$00404.}

\ignore{
\EventEditors{Xavier Goaoc and Michael Kerber}
\EventNoEds{2}
\EventLongTitle{38th International Symposium on Computational Geometry (SoCG 2022)}
\EventShortTitle{SoCG 2022}
\EventAcronym{SoCG}
\EventYear{2022}
\EventDate{June 7--10, 2022}
\EventLocation{Berlin, Germany}
\EventLogo{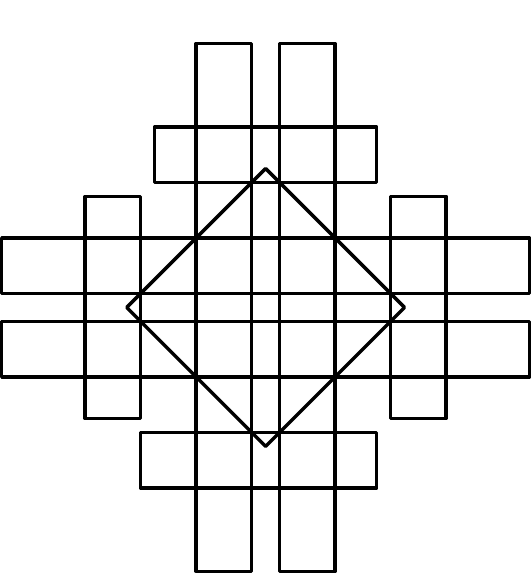}
\SeriesVolume{224}
\ArticleNo{XX}  % <-- This will be filled in by the typesetters
}
\begin{document}

\maketitle

\begin{abstract}

Semialgebraic range searching, 
arguably the most general version of range searching,
is a fundamental problem in computational geometry.
In the problem, we are to preprocess a set of points in $\bR^D$ such that the subset of points inside
a semialgebraic region described by a constant number of polynomial inequalities of degree $\Delta$ can be
found efficiently.

Relatively recently, several major advances were made on this problem.
Using algebraic techniques, ``near-linear space'' data structures~\cite{agarwal2013on, matousek2015multilevel}
with almost optimal query time of $Q(n)=O(n^{1-1/D+o(1)})$ were obtained.
For ``fast query'' data structures (i.e., when $Q(n)=n^{o(1)}$), it was conjectured
that a similar improvement is possible,
i.e., it is possible to achieve space $S(n) = O(n^{D+o(1)})$.
The conjecture was refuted very recently by Afshani and Cheng~\cite{afshani2021lower}.
In the plane, i.e., $D=2$, they proved that $S(n) = \Omega(n^{\Delta+1 - o(1)}/Q(n)^{(\Delta+3)\Delta/2})$
which shows
$\Omega(n^{\Delta+1-o(1)})$ space is needed for $Q(n) = n^{o(1)}$.
While this refutes the conjecture,
it still leaves a number of unresolved issues:
the lower bound only works in 2D and for fast queries, 
and neither the exponent of $n$ or $Q(n)$ seem to be tight even for $D=2$, as
the best known upper bounds have $S(n) = O(n^{\bm+o(1)}/Q(n)^{(\bm-1)D/(D-1)})$
where $\bm=\binom{D+\D}{D}-1 = \Omega(\Delta^D)$ is the maximum number of parameters to
define a monic degree-$\D$ $D$-variate polynomial, for any constant dimension $D$ and degree $\D$.

In this paper, we  resolve two of the issues:
we prove a lower bound in $D$-dimensions, for constant $D$, and 
show that when the query time is $n^{o(1)}+O(k)$,
the space usage is $\Omega(n^{\bm-o(1)})$, which
almost matches the $\tilde O(n^{\bm})$ upper bound 
and essentially closes the problem for the fast-query case, as far as the exponent of $n$ 
is considered in the pointer machine model.
When considering the exponent of $Q(n)$, 
we show that the analysis in~\cite{afshani2021lower} is tight for $D=2$,
by presenting matching upper bounds for uniform random point sets.
This shows either the existing upper bounds can be improved or to obtain
better lower bounds a new fundamentally different input set needs to be constructed. 

\end{abstract}

\section{Introduction}

In the classical semialgebraic range searching problem, we are to preprocess a set of $n$ points in $\bR^D$ such that the subset of points inside
a semialgebraic region, described by a constant number of polynomial inequalities of degree $\Delta$ can be
found efficiently.
Recently, two major advances were made on this problem.
First, in 2019, Agarwal~\etal~\cite{agarwal2021efficient} showed for polylogarithmic query time,
it is possible to build a data structure of size $\tilde O(n^\beta)$ space\footnote{
	$\tilde\Omega(\cdot),\tilde O(\cdot),\tilde\Theta(\cdot)$ notations hide $\log^{o(1)}n$ factors;
	$\domega(\cdot),\dO(\cdot),\dtheta(\cdot)$ notations hide $n^{o(1)}$ factors.
},
where $\beta$ is the number of parameters needed to specify a query polynomial.
For example, for $D=2$, a query polynomial is in the form of $\sum_{i+j\le\D}a_{ij}x^iy^j\le0$
where $a_{ij}$'s are specified at the query time,
and when $\D=4$,
$\beta$ can be as large as $14$ (technically, there are 15 coefficients but one
coefficient can always be normalized to be 1).
In this case,
a major conjecture was that if this space bound
could be improved to $\tilde O(n^D)$
(e.g., for $\Delta=4$, from $\tilde O(n^{14})$
to $\tilde O(n^2)$).
Very recently, Afshani and Cheng~\cite{afshani2021lower}
refuted this conjecture by showing a $\domega(n^{\D+1})$ lower bound.
However, there are two major limitations of their lower bound.
First, their lower bound only works in $\bR^2$,
while the upper bound in~\cite{agarwal2021efficient} holds for all dimensions. 
Second, their lower bound only works for queries of form $y-\sum_{i=0}^{\D}x^i\le0$
and thus their lower bound does not give a satisfactory answer to the problem
in the general case.
For example, for $D=2,\D=4$,
they show a $\domega(n^5)$ lower bound
whereas the current best upper bound is $\tilde O(n^{14})$.
In general, their space lower bound is at most $\domega(n^{\D+1})$
while the upper bound of~\cite{agarwal2021efficient} can be $\tilde O(n^{\Theta(\D^2)})$,
which leaves an unsolved wide gap, even for $D=2$.
Another problem brought by~\cite{agarwal2021efficient} is the space-time tradeoff.
When restricted to queries of the form  $y-\sum_{i=0}^{\D}x^i\le0$,
the current upper bound tradeoff is $S(n)=\tilde O(n^{\Delta+1}/Q(n)^{2\Delta})$~\cite{matousek2015multilevel, agarwal2021efficient}
while the lower bound in~\cite{afshani2021lower} is $S(n)=\domega(n^{\Delta+1}/Q(n)^{(\Delta+3)\Delta/2})$.
Even for $\D=2$, we observe a discrepancy between an $S(n)=\tilde O(n^3/Q(n)^4)$ upper
and an $S(n)=\domega(n^3/Q(n)^5)$ lower bound.

Here, we make progress in both lower and upper bound directions.
We give a general lower bound in $D$ dimensions
that is tight for all possible values of $\beta$.
Our lower bound attains the maximum possible $\beta$ value $\bm_{D,\D}=\binom{D+\D}{D}-1$,
e.g., $\domega(n^{14})$ for $D=2,\D=4$.
Thus, our lower bounds almost completely settle the general case of the problem for the fast-query case,
as far as the exponent of $n$ is concerned.
This improvement is quite non-trivial and requires significant new insights that are not avaiable in~\cite{afshani2021lower}.
For the upper bound, we present a matching space-time tradeoff
for the two problems studied in~\cite{afshani2021lower}
for uniform random point sets.
This shows their lower bound analysis is tight. 
Since for most range searching problems,
a uniform random input instance is the hardest one,
our results show that current upper bound based on the classical method 
might not be optimal.
We develop a set of new ideas for our results
which we believe are important for further investigation
of this problem.

% I deleted some simplex results etc.
% If you think we need to mention them
% they are commented in the ignore part
\subsection{Background}

In range searching,
the input is a set of points in $\bR^D$
for a fixed constant $D$.
The goal is to build a structure
such that for a query range,
we can report or find the points in the range efficiently.
This is a fundamental problem in computational geometry
with many practical uses in e.g., databases and GIS systems.
For more information, see surveys by Agarwal~\cite{handbookdcg2018}
or Matou\v sek~\cite{matousek1994geometric}.
We focus on a fundamental case of the problem
where the ranges are semialgebraic sets of constant complexity
which are defined by intersection/union/complementation 
of $O(1)$ polynomial inequalities of constant degree at most $\D$ in $\bR^D$.

The study of this problem dates back to at least 35 years ago~\cite{yao1985ageneral}.
A linear space and $O(n^{1-1/D+o(1)})$ query time structure
is given by Agarwal, Matou\v sek, and Sharir~\cite{agarwal2013on},
due to the recent ``polynomial method'' breakthrough~\cite{guth2015on}.
However, it is not entirely clear what happens to the ``fast-query'' case:
if we insist on polylogarithmic query time,
what is the smallest possible space usage?
Early on, some believed that the number of parameters
plays an important role and thus $\tilde O(n^{\beta})$ space 
could be a reasonable conjecture~\cite{matousek1994geometric}, but
such a data structure was not found until 2019~\cite{agarwal2021efficient}.
However, after the ``polynomial method'' revolution, and specifically
after the breakthrough result of Agarwal, Matou\v sek and Sharir~\cite{agarwal2013on},
it could also be reasonably conjectured that $\tilde{O}(n^{D})$ could also be the right bound.
However, this was refuted recently by 
Afshani and Cheng~\cite{afshani2021lower} who
showed that in 2D, and for polynomials for the form $y-\sum_{i=0}^{\D}x^i\le0$, there exists
an $\domega(n^{\Delta+1})$ space lower bound for data structures with query time $\dO(1)$. 
However, this lower bound does not go far enough, even in 2D,
where a semialgebraic range can be specified by 
bivariate monic polynomial inequalities\footnote{
	We define that a $D$-variate polynomial $P(X_1,X_2,\cdots,X_D)$
	is monic if the coefficient of $X_2^{\D}$ is $-1$.
} 
of form $\sum_{i,j:i+j\le\D}a_{ij}x^iy^j\le0$ with $a_{\D0}=-1$.
In this case, $\beta$ can be as large as $\bm_{2,\D}=\binom{\D+2}{2}-1=\Theta(\D^2)$, 
and much larger than $\D+1$ even for moderate $\D$ (e.g., for $\Delta=4$, ``5'' versus ``14'', for
$\Delta=5$, ``6'' versus ``20'' and so on).
Another main weakness is that their lower bound is only in 2D,
but the upper bound~\cite{agarwal2021efficient} works in arbitrary dimensions.

The correct upper bound tradeoff seems to be even more mysterious.
Typically, the tradeoff is obtained by combining the linear space and the polylogarithmic query time solutions.
For simplex range searching (i.e., when $\Delta=1$), the tradeoff is
$S(n) = \tilde O(n^D/Q(n)^D)$~\cite{matousek1993range}, which is a natural looking bound and it is also known to be optimal. 
The tradeoff bound becomes very mysterious for semialgebraic range searching. 
For example, for $D=2$ and when restricted to queries of the form $y-\sum_{i=0}^{\D}x^i\le0$,
combining the existing solutions yields the bound $S(n)= \tilde O(n^{\Delta+1}/Q(n)^{2\Delta})$
whereas the known lower bound~\cite{afshani2021lower} is $S(n)= \domega (n^{\Delta+1}/Q(n)^{(\Delta+3)\Delta/2})$.
One possible reason for this gap
is that the lower bound construction is based on a uniform random point set,
while in practice, the input can be pathological.
But in general the uniform random point set assumption is not too restrictive for range searching problems.
Almost all known lower bounds rely on this assumption: 
e.g., half-space range searching~\cite{bronnimann1993how,arya2006on,arya2012tight},
orthogonal range searching~\cite{chazelle1990loweri,chazelle1990lowerii,afshani2019anew}, 
simplex range searching~\cite{chazelle1989lower,chazelle1996simplex,afshani2012improved}.

\subsection{Our Results}
Our results consist of two parts.
First, we study a problem that we call ``the general polynomial slab range reporting''.
Formally, let $P(X)$ be a monic $D$-variate polynomial of degree at most $\D$,
a general polynomial slab is defined to be the region between $P(X)=0$ and $P(X)=w$
for some parameter $w$ specified at the query time.
Unlike~\cite{afshani2021lower},
our construction can reach the maximum possible parameter number $\bm_{D,\D}$.
For simplicity, we use $\bm$ instead of $\bm_{D,\D}$ when the context is clear.
We give a space-time tradeoff lower bound of $S(n)=\domega(n^{\bm}/Q(n)^{\Theta((\D^2+D\D)\bm}))$,
which is (almost) tight when $Q(n)=n^{o(1)}$.

For the second part, we present
data structures that match the lower bounds studied in the work by Afshani and Cheng~\cite{afshani2021lower}.
We show that their lower bounds for 2D polynomial slabs and 2D annuli are tight
for uniform random point sets.
Our bound shows that current tradeoff given by the classical method of combining 
extreme solutions~\cite{matousek2015multilevel, agarwal2021efficient} might not be tight.
We shred some lights on the upper bound tradeoff and
develop some ideas which could be used to tackle the problem.
Our results are summarized in Table~\ref{tab:results}.

{\small
\begin{table*}[h]
\centering
\setlength\extrarowheight{2.5pt}
\caption{Our Results (marked by $^*$). Our upper bounds are for uniform random point sets.}
\label{tab:results}
\begin{tabular}{ | m{3.65cm} | m{4.5cm}| m{4.1cm} |} 
\hline
\bf{Query Types}& \bf{Lower Bound} & \bf{Upper Bound} \\
\hline
	General Polynomial Slabs $\left(\bm=\bm_{D,\D}=\binom{D+\D}{D}-1\right)$ & 
	$S(n)=\domega\left(\frac{n^{\bm}}{Q(n)^{\Theta(\bm)}}\right)^*$ & 
	$S(n)=\tilde{O}\left(\frac{n^{\bm}}{Q(n)^{\Theta(\bm)}}\right)$~\cite{matousek2015multilevel, agarwal2021efficient} \\
	\bf{When} $\boldsymbol{Q(n)=\dO(1)}$ & 
	$\boldsymbol{S(n)=\domega\left(n^{\bm}\right)^*}$ & 
	$\boldsymbol{S(n)=\tilde{O}\left(n^{\bm}\right)}$~\cite{matousek2015multilevel, agarwal2021efficient} \\
\hline
\hline
	2D Semialgebraic Sets $\left(\bm=\bm_{2,\D}=\binom{2+\D}{2}-1\right)$ & 
	$S(n)=\domega
	\left(
		\frac{n^{\bm}}{Q(n)^{\bm+\bm^2(\bm-1)-1}}
	\right)^*$ & 
	
	$S(n)=\tilde{O} 
	\left(
		\frac{ n^{\bm} }{ Q(n)^{2\bm-2} }
	\right)$~\cite{matousek2015multilevel, agarwal2021efficient} 
	$\boldsymbol{S(n)=\tilde{O} 
	\left(
		\frac{ n^{\bm} }{ Q(n)^{3\bm-4} }
	\right)^*}$\\
\hline
	\bf{2D Polynomial Slabs} & 
	$\boldsymbol{S(n)=\domega
	\left(
		\frac{n^{\D+1}}{Q(n)^{(\D+3)\D/2}}
	\right)}$~\cite{afshani2021lower} & 
	
	$S(n)=\tilde{O} 
	\left(
		\frac{ n^{\D+1} }{ Q(n)^{2\D}}
	\right)$~\cite{matousek2015multilevel, agarwal2021efficient} 
	$\boldsymbol{S(n)=\tilde{O} 
	\left(
		\frac{ n^{\D+1} }{ Q(n)^{(\D+3)\D/2} }
	\right)^*}$\\
\hline
	\bf{2D Annuli} & 
	$\boldsymbol{S(n)=\domega
	\left(
		\frac{n^{3}}{Q(n)^{5}}
	\right)}$~\cite{afshani2021lower} & 
	
	$S(n)=\tilde{O} 
	\left(
		\frac{n^{3} }{ Q(n)^{4}}
	\right)$~\cite{matousek2015multilevel, agarwal2021efficient} 
	$\boldsymbol{S(n)=\tilde{O} 
	\left(
		\frac{n^{3} }{ Q(n)^{5}}
	\right)^*}$\\
\hline
\end{tabular}
\end{table*}
}

\subsection{Technical Contributions}
Compared to the previous lower bound in~\cite{afshani2021lower},
we need to wrestle with many complications that stem from
the algebraic geometry nature of the problem.
In Section~\ref{sec:genpolyslablb}, we cover them in greater detail,
but briefly speaking, the technical heart of the results
in~\cite{afshani2021lower} is that ``two univariate polynomials
$P_1(x)$ and $P_2(x)$ that have sufficiently different leading coefficients, 
cannot pass close to each other for too long.
However, this claim is not true for even bivariate polynomials,
since $P_1(x,y)$ and $P_2(x,y)$ could have infinitely many roots in common
and thus we can have $P_1(x,y)-P_2(x,y)=0$ in an unbounded region of $\bR^2$.
Overcoming this requires significant innovations. 

\section{Preliminaries}

In this section, we introduce some tools we will use in this paper.
We will mainly use the lower bound tools used in~\cite{afshani2021lower}.
For more detailed introduction,
we refer the readers to~\cite{afshani2021lower}.

\subsection{A Geometric Lower Bound Framework}

We present a lower bound framework in the pointer machine model of computation.
It is a streamlined version of the framework by 
Chazelle~\cite{chazelle1990loweri} and Chazelle and Rosenberg~\cite{chazelle1996simplex}.
In essence, this is an encapsulation of the way
the framework is used 
in~\cite{afshani2021lower}.

In a nutshell, in the pointer machine model,
the memory is represented as a directed graph 
where each node can store one point and it has two pointers
to two other nodes.
Given a query, starting from a special ``root'' node, 
the algorithm explores a subgraph that contains
all the input points to report.
The size of the explored subgraph is the query time.

Intuitively, for range reporting,
to answer a query fast,
we need to store its output points close to each other.
If each query range contains many points to report
and two ranges share very few points,
some points must be stored multiple times,
thus the total space usage must be big.
% to do: rewrite the intuition
We present the framework,
and refer the readers to the Appendix~\ref{sec:proof-rrfw} for the proof.

\begin{restatable}{theorem}{rrfw}\label{thm:rrfw}
	Suppose the $D$-dimensional geometric range reporting problems
	admit an $S(n)$ space $Q(n)+O(k)$ query time data structure,
	where $n$ is the input size and $k$ is the output size.
	Let $\mu^D(\cdot)$ denote the $D$-dimensional Lebesgue measure.
	Assume we can find $m=n^{c}$ ranges $\rR_1, \rR_2, \cdots, \rR_m$ in a
	$D$-dimensional cube $\C^D$ of side length $|l|$ for some constant $c$ such that
	(i) $\forall i=1,2,\cdots,m, \mu^D(\rR_i\cap\C^D)\ge 4c|l|^DQ(n)/n$; and
	(ii) $\mu^D(\rR_i\cap\rR_j)=O(|l|^D/(n2^{\sqrt{\log n}}))$ for all $i\neq j$.
	Then, we have $S(n)=\domega(mQ(n))$.
\end{restatable}

\subsection{A Lemma for Polynomials}
Given a univariate polynomial and some positive value $w$, 
the following lemma from~\cite{afshani2021lower} upper bounds
the length of the interval within which the absolute value of the polynomial is no more than $w$.
We will use this lemma as a building block for some of our proofs.
\begin{lemma}[Afshani and Cheng~\cite{afshani2021lower}]
\label{lem:polylem}
	Given a degree-$\Delta$ univariate polynomial $P(x)=\sum_{i=0}^{\Delta}a_ix^i$ where $|a_{\Delta}|>0$
	and $\Delta>0$.
	Let $w$ be any positive value.
	If $|P(x)| \le w$ for all $x \in [x_0, x_0+t]$ for some parameter $x_0$, then
	$t =O((w/|a_{\Delta}|)^{1/\Delta})$.
\end{lemma}

\subsection{Useful Properties about Matrices}
In this section, we recall some useful properties about matrices.
We first recall some properties of the determinant of matrices.
One important property is that the determinant is mutilinear:

\begin{lemma}
\label{lem:linmat}
	Let $A=\begin{bmatrix}\ba_1 & \cdots & \ba_n \end{bmatrix}$ be a $n\times n$ matrix
	where $\ba_i$'s are vectors in $\bR^n$. Suppose $\ba_j=r \cdot \bw + \bv$ for some $r\in\bR$
	and $\bw,\bv\in\bR^n$, then the determinant of $A$, denoted $\det(A)$, is
	\[
	\begin{aligned}
		\det(A)
		&=
		\det\left(
			\begin{bmatrix}
				\ba_1 & \cdots & \ba_{j-1} & \ba_j & \ba_{j+1} & \cdots & \ba_n
			\end{bmatrix}
		\right)\\
		&=
		r\cdot \det\left(
			\begin{bmatrix}
				\ba_1 & \cdots & \ba_{j-1} & \bw & \ba_{j+1} & \cdots & \ba_n
			\end{bmatrix}
		\right)
		+
		\det\left(
			\begin{bmatrix}
				\ba_1 & \cdots & \ba_{j-1} & \bv & \ba_{j+1} & \cdots & \ba_n
			\end{bmatrix}
		\right).			
	\end{aligned}
	\]
\end{lemma}

One of the special types of matrices we will use is the Vandermonde matrix
which is a square matrix
where the terms in each row form a geometric series, i.e., $V_{ij}=x_i^{j-1}$ for all indices $i$ and $j$.
The determinant of such a  matrix is 
$
	\det(V) = \prod_{1\le i < j \le n}(x_j-x_i).
$

Given an $n$-tuple $\lambda=(\lambda_1,\lambda_2, \cdots, \lambda_n)$
where $\lambda_1\ge\lambda_2\ge\cdots\ge\lambda_n\ge0$,
we can define a generalized Vandermonde matrix $V^*$ defined by $\lambda$, 
where $V^*_{ij}=x_i^{\lambda_{n-j+1}+j-1}$.
The determinant of $V^*$ is known to be the product of
the determinant of the induced Vandermonde matrix $V_{V^*}$ with $V_{ij}=x_i^{j-1}$ and the Schur polynomial 
$s_{\lambda}(x_1,x_2,\cdots,x_n)=\sum_Tx_1^{t_1}\cdots x_n^{t_n}$,
where the summation is over all semistandard Young tableaux~\cite{young1901} $T$ of shape $\lambda$.
The exponents $t_1,t_2,\cdots,t_n$ are all nonnegative numbers.
The following lemma bounds the determinant of a generalized Vandermonde matrix.

\begin{lemma}
\label{lem:genvanddet}
	Let $V^*$ be a generalized Vandermonde matrix defined by $\lambda=(\lambda_1,\lambda_2,\cdots,\lambda_n)$
	where $\lambda_1\ge\lambda_2\ge\cdots\ge\lambda_n\ge0$.
%  	If $n=\Theta(1)$, and $V^*_{ij} = \Theta(1)$, then $\det(V^*)=\Theta(\det(V_{V^*}))$,
	If $n,\lambda_1=\Theta(1)$, and for all $i$, $x_i = \Theta(1)$, then $\det(V^*)=\Theta(\det(V_{V^*}))$,
	where $V_{V^*}$ is the induced Vandermonde matrix with $V_{ij}=x_i^{j-1}$.
\end{lemma}

\section{Lower Bound for Range Reporting with General Polynomial Slabs}
\label{sec:genpolyslablb}
In this section, we prove our main lower bound for general polynomial slabs.

\begin{definition}
	A general polynomial slab in $\bR^D$ is a triple $(P, a, b)$ 
	where $P\in\bR[X]$ is a degree-$\D$ $D$-variate polynomial and $a, b$ are two real numbers such that $a<b$.
	A general polynomial slab is defined as $\{X\in\bR^D: a \le P(X) \le b\}$.
	Note that due to rescaling, we can assume that the polynomial is monic.
\end{definition}

Before presenting our results, 
we first describe the technical challenges of this problem.
We explain why the construction used in~\cite{afshani2021lower}
cannot be generalized in an obvious way
and give some intuition behind our lower bound construction.

\subsection{Technical Challenges}
Our goal is a lower bound of the form $\domega(n^{\bm}/Q(n)^{\Theta(\bm)})$.
To illustrate the challenges,
consider the case $D=2$ and the unit square $\U=\U^2=[0,1]\times[0,1]$.
To use Theorem~\ref{thm:rrfw},
we need to generate about $\domega(n^{\bm})$ polynomial slabs
such that each slab should have width approximately $\Omega(Q(n)/n)$, 
and any two slabs should intersect with area approximately $O(1/n)$.
Intuitively, this means two slabs cannot intersect
over an interval of length $\Omega(1/Q(n))$.
    
In Lemma~\ref{lem:polylem},
for univariate polynomials,
the observation behind their construction is that 
when the leading coefficients of two polynomials differ by a large number,
the length of the interval in which two polynomials are close to each other is small.
However, when we consider general bivariate polynomials in $\bR^2$,
this observation is no longer true.
For example, 
consider $P_1(x,y)=(x+1)(1000x^2+y)$ and $P_2(x,y)=(x+1)(x^2+1000y)$.
The leading coefficients are $1000$ and $1$ respectively,
but since $P_1,P_2$ have a common factor $(x+1)$, 
their zero sets have a common line.
Thus any slab of width $Q(n)/n$ generated for these two polynomial 
will have infinite intersection area,
which is too large to be useful.

At first glance,
it might seem that this problem can be fixed by picking the polynomials
randomly, e.g., each coefficient is picked independently
and uniformly from the interval $[0,1]$,
as a random polynomial in two or more variables is irreducible with probability $1$.
Unfortunately, this does not work either but for some very nontrivial reasons.
To see this,
consider picking
coefficients uniformly
at random from range $[0,1]$
for bivariate polynomials $P(x,y)=\sum_{i+j\le\D}a_{ij}x^iy^j$.
The probability of pick a polynomial
with $0 \le a_{0j} \le \frac{1}{n}$ for all $a_{0j}$
is $\frac{1}{n^{\Delta+1}}$.
For such polynomials,
$0\le P(0,y) \le \frac{\Delta+1}{n}$ for $y\in[0,1]$.
Suppose we sampled two such polynomials, then the two slabs generated using them
will contain $x=0$ for $y\in[0,1]$,
meaning, the two slabs will have too large of an area ($\Omega(Q(n)/n)$) in common, so we cannot have that. 
Unfortunately, if we sample more than $n^{\Delta+1}$ polynomials,
this will happen with probability close to one, and there seems to be no easy fix.
A deeper insight into the issue is given below.

Map a polynomial $\sum_{i+j\le\D} a_{ij}x^iy^j$ to the point $(a_{00}, a_{01}, \cdots, a_{\D0})$ in $\bR^{\bm}$.
The above randomized construction corresponds to picking
a random point from the unit cube $\U$ in $\bR^{\bm}$.
Now consider the subset $\Gamma$ of $\bR^{\bm}$ that
corresponds to reducible polynomials.
The issue is that $\Gamma$ intersects $\U$
and thus we will sample polynomials that are close to reducible
polynomials, e.g., a sampled polynomial with $a_{0j}=0\in[0,\frac{1}{n}]$
is close to the reducible polynomial with $a_{0j}=0$.
Pick a large enough sample and two points will lie close to the same reducible polynomial
and thus they will produce a ``large'' overlap in the construction. 
Our main insight is that there exists a point $\bp$ in $\U$ that has a ``fixed'' (i.e., constant)
distance to $\Gamma$; thus, we can consider a neighborhood around $\bp$ and sample our polynomials
from there.
However, more technical challenges need to be overcome to even make this idea work but it
turns out, we can simply pick our polynomials from a grid constructed in the small enough neighborhood
of some such point $\bp$ in $\bR^{\bm}$.

\subsection{A Geometric Lemma}
In this section, we show a geometric lemma
which we will use to establish our lower bound.
In a nutshell, given two monic $D$-variate polynomials $P_1, P_2$
and a point $p=(p_2,p_3,\cdots,p_D)\in\bR^{D-1}$
in the $(D-1)$-dimensional subspace
perpendicular to the $X_1$-axis,
we define the distance between $Z(P_1)$\footnote{$Z(P)$ denotes the zero set of polynomial $P$.} and $Z(P_2)$ along the $X_1$-axis
at point $p$ to be $|a-b|$, where 
$(a,p_2,\cdots,p_D)\in Z(P_1)$
and $(b,p_2,\cdots,p_D)\in Z(P_2)$.
In general, this distance is not well-defined
as there could be multiple $a$ and $b$'s satisfying the definition.
But we can show that for a specific set of polynomials,
$a,b$ can be made unique and thus the distance is well-defined.
For $P_1,P_2$ with ``sufficiently different'' coefficients,
we present a lemma which upper bounds
the $(D-1)$-measure of the set of points $p$
at which the distance between $Z(P_1)$ and $Z(P_2)$ is ``small''.
Intuitively, this can be viewed as a generalization of Lemma~\ref{lem:polylem}.
We first prove the lemma in 2D for bivariate polynomials,
and then extend the result to higher dimensions.

First, we define the notations we will use for general $D$-variate polynomials.
\begin{definition}
\label{def:simpdef}
	Let $I^D\subseteq\{(i_1,i_2,\cdots,i_D)\in\mathbb{N}^D\}$\footnote{In this paper, $\mathbb{N}=\{0, 1,2,\cdots\}$.}, $D\ge1$,
	be a set of $D$-tuples where each tuple consists of nonnegative integers.
	We call $I^D$ an index set (of dimension $D$).
	Let $X^D=(X_1,X_2,\cdots,X_D)$ be a $D$-tuple of indeterminates.
	When the context is clear, we use $X$ for simplicity.
	Given an index set $I^D$, we define
	\[
		P(X) = \sum_{i\in I^D}A_iX^i,
	\]
	where $A_i\in\bR$ is the coefficient of $X^i$ and $X^i=X_1^{i_1}X_2^{i_2}\cdots X_D^{i_D}$,
	to be a $D$-variate polynomial.
	For any $i\in I^D$, we define $\sigma(i)=\sum_{j=1}^Di_j$.
	Let $\D$ be the maximum $\sigma(i)$ with $A_i\neq 0$, and
	we say $P$ is a degree-$\D$ polynomial.
	Given a $D$-tuple $T$, we use $T_{:j}$ to denote a $j$-tuple by taking only the first $j$ components of $T$.
	Also, we use notation $T_j$ to specify the $j$-th component of $T$.
	Conversely, given a $(D-1)$-tuple $t$ and a value $v$,
	we define $t\oplus v$ to be the $D$-tuple formed by appending $v$ to the end of $t$.
\end{definition}

We will consider polynomials of form
\[
	P(X)=X_1-X_2^{\D}+\sum_{i\in I^D}A_iX^i,
\]
where $0\le A_{ij} = O(\epsilon)=o(1)$ for all $\sigma(i)\le\D$
except that $A_i=0$ for $i=(0,\D,0,\cdots,0)$.
Intuitively, these are monic polynomials
packed closely in the neighborhood of  $P(X)=X_1-X_2^D$.
For simplicity, we call them ``packed'' polynomials.
We will prove a property for packed polynomials
that are ``sufficiently distant''. More precisely,

\begin{definition}
\label{def:distant}
	Given two distinct packed degree-$\D$ $D$-variate polynomials $P_1,P_2$,
	we say $P_1,P_2$ are ``distant'' if each coefficient of $P_1-P_2$ 
	has absolute value at least $\xi_D=\delta\tau^{\bbb}(\eta\tau)^{(D-2)\Delta}>0$ if not zero 
	for parameters $\delta, \eta, \tau > 0$ and $\eta\tau=O((1/\epsilon)^{1/\bbb})$,
	where $\bbb=\binom{\bb}{2}$
	and $\bb=\bm_{2,\D}$ is the maximum number of coefficients needed 
	to define a monic degree-$\Delta$ bivariate polynomial.
\end{definition}

We will use the following simple geometric observation.
See Appendix~\ref{sec:proof-func} for the proof.
\begin{restatable}{observation}{func}\label{obs:func}
	Let $P$ be a packed $D$-variate polynomial and 
	$a=(a_1,a_2,\cdots,a_D)\in Z(P)$.
	If $a_i\in[1,2]$ for all $i=2,3,\cdots,D$,
	then there exists a unique $a_1$ such that $0<a_1=O(1)$.
\end{restatable}

With this observation, we can define the distance between the zero sets of two polynomials
along the $X_1$-axis at a point in $[1,2]^{D-1}$ of the subspace
perpendicular to the $X_1$ axis.

\begin{definition}
	Given two packed polynomials $P_1, P_2$
	and a point $p=(p_2,p_3,\cdots,p_D)\in[1,2]^{D-1}$,
	we define the distance between $Z(P_1)$ and $Z(P_2)$ at point $p$, denoted by $\pi(Z(P_1), Z(P_2), p)$, to be
	$|a-b|$ s.t. $a,b>0$, and $(a,p_2,p_3,\cdots,P_D)\in Z(P_1)$
	and $(b,p_2,p_3,\cdots,P_D)\in Z(P_2)$.
\end{definition}

Now we show a generalization of Lemma~\ref{lem:polylem} to distant bivariate polynomials in 2D.

\begin{lemma}
\label{lem:intlen}
	Let $P_1,P_2$ be two distinct distant bivariate polynomials.
	Let $I=\{y:\pi(Z(P_1),Z(P_2),y)=O(w) \land y \in[1,2] \}$, where $w=\delta/\eta^{\bbb}=o(1)$.
	Then $|I|=O(\frac{1}{\eta\tau})$.
\end{lemma}

\begin{proof}
	We prove it by contradiction.
	The idea is that if the claim does not hold,
	then we can ``tweak'' the coefficients of $P_2$
	by a small amount such that the tweaked polynomial and $P_1$
	have $\bb$ common roots.
	Next, we show this implies that the tweaked polynomial is equivalent to $P_1$.
	Finally we reach a contradiction by noting that by assumption 
	at least one of the coefficients of $P_1$ and $P_2$ is not close.
	Let $P_1(x,y)=x-y^{\Delta}+\sum_{i=0}^{\Delta}\sum_{j=0}^{\Delta-i}a_{ij}x^iy^j$ and
	$P_2(x,y)=x-y^{\Delta}+\sum_{i=0}^{\Delta}\sum_{j=0}^{\Delta-i}b_{ij}x^iy^j$
	where by definition all $a_{ij}$'s and $b_{ij}$'s are $O(\epsilon)$.
  	Suppose for the sake of contradiction that $|I|=\omega(\frac{1}{\eta\tau})$.
	We pick $\bb$ values $y_1,y_2,\cdots,y_{\bb}$ in $I$ s.t. $|y_i-y_j|\ge |I|/\bb$ for all $i\neq j$.
	Let $x_1,x_2,\cdots,x_{\bb}$ be the corresponding values s.t.
	$(x_k,y_k)\in Z(P_1)$ in the first quadrant, i.e., $P_1(x_k,y_k)=0$ for $k=1,2,\cdots,\bb$.
	Note that
	\[
		P_1(x_k,y_k)=0\equiv x_k-y_k^{\Delta}+\sum_{i=0}^{\Delta}\sum_{j=0}^{\Delta-i}a_{ij}x_k^iy_k^j=0\implies x_k=y_k^{\Delta}-O(\epsilon),
	\]
	since $a_{ij}=O(\epsilon)$ and $x_k, y_k = O(1)$ by Observation~\ref{obs:func}.
	Since $\pi(Z(P_1),Z(P_2), y_k)=O(w)$ for all $y_k\in I$,
  	let $(x_k+\D x_k,y_k)$ be the points on $Z(P_2)$,
	we have $P_{2}(x_k + \D x_k, y_k)=P_{2}(x_k, y_k)+\Theta(\D x_k)=0$.
	Since $|\D x_k|=O(w)$, $P_{2}(x_k,y_k)=\bg_k$ for some $|\bg_k|=O(w)$.
  	We would like to show that we can ``tweak'' every coefficient $b_{ij}$ of $P_{2}(x,y)$ by some value $\bd_{ij}$,
  	to turn $P_{2}$ into a polynomial $Q$
  	s.t. $Q(x_k, y_k)=0, \forall k=1,2,\cdots,\bb$. 
  	If so, for every pair $(x_k,y_k)$,
	\[
		\begin{aligned}
			Q(x_k,y_k)
				&=x_k-y_k^{\Delta}+\sum_{i=0}^{\Delta}\sum_{j=0}^{\Delta-i}(b_{ij}+\bd_{ij})x_k^iy_k^j\\
				&=P_{2}(x_k,y_k)+\sum_{i=0}^{\Delta}\sum_{j=0}^{\Delta-i}\bd_{ij}x_k^iy_k^j\\
				&=\bg_k+\sum_{i=0}^{\Delta}\sum_{j=0}^{\Delta-i}\bd_{ij}(y_k^{\Delta}-O(\epsilon))^iy_k^j\\
				&=\bg_k+\sum_{i=0}^{\Delta}\sum_{j=0}^{\Delta-i}\bd_{ij}(y_k^{i\Delta}-O(\epsilon))y_k^j,\\
		\end{aligned}
	\]
	where the last equality follows from $\epsilon = o(1)$ and $1\le y_k\le 2$.
	So to find $\bd_{ij}$'s and to be able to tweak $P_2(x,y)$, we need to solve the following linear system
	\begin{align*}
		\begin{bmatrix}
		1 & y_1 & y_1^2 & \cdots & y_1^{\Delta-1} & y_1^{\Delta}-O(\epsilon) & \cdots & y_1^{\Delta^2}-O(\epsilon)\\
		1 & y_2 & y_2^2 & \cdots & y_2^{\Delta-1} & y_2^{\Delta }-O(\epsilon) & \cdots & y_2^{\Delta^2}-O(\epsilon)\\
		\vdots & \vdots & \vdots & \ddots & \vdots & \vdots & \ddots & \vdots\\
		1 & y_{\bb} & y_{\bb}^2 & \cdots & y_{\bb}^{\Delta-1} & y_{\bb}^{\Delta}-O(\epsilon) & \cdots & y_{\bb}^{\Delta^2}-O(\epsilon)\\
		\end{bmatrix}
		\cdot
		\begin{bmatrix}
		\bd_{00} \\ \bd_{01} \\ \vdots \\ \bd_{\Delta0}
		\end{bmatrix}
		=
		\begin{bmatrix}
		-\bg_1 \\ -\bg_2 \\ \vdots \\ -\bg_{\bb}
		\end{bmatrix},		
	\end{align*}
	where the exponents of $y_k$ are generated by $i\D + j$ for $i, j \in \{0,1,2,\cdots,\Delta\}$, $j\neq\D$, and $i+j\le \Delta$.
	Let us call the above linear system $A\cdot \bd = \bg$. 
	
	By Lemma~\ref{lem:linmat}, $\det(A)=\det(A^*)+\sum_{l=1}^{\Theta(1)}\det(A_l)$,
	where $A^*$ is a generalized Vandermonde matrix
	defined by an $\bb$-tuple $\lambda=(\D^2-\bb,\dots,0)$,
	and each $A_l$ is a matrix with some columns being $O(\epsilon)$.
	Since $\bb=\binom{2+\D}{2}-1$ is $\Theta(1)$, by Lemma~\ref{lem:genvanddet},
	we can bound $\det(A^*)$ by $\Theta(\det(V_{A^*}))$, where $V_{A^*}$ is the induced Vandermonde matrix.
	Since $|y_i-y_j|=\Omega(|I|)$ for $i\neq j$, $\det(V_{A^*})=\prod_{1\le i < j\le\bb}(y_j-y_i))=\Omega(|I|^{\bbb})$.
	On the other hand, for every matrix $A_l$, 
	there is at least one column where the magnitude of all the entries is $O(\epsilon)$.
	Since all other entries are bounded by $O(1)$,
	by the Leibniz formula for determinants, $|\det(A_l)|=O(\epsilon)=O((\frac{1}{\eta\tau})^{\bbb})$.
	Since $|I|^{\bbb}=\omega((\frac{1}{\eta\tau})^{\bbb})$,
	we can bound $|\det(A)|=\Omega(|I|^{\bbb})$ and in particular
  	$|\det(A)|\neq 0$ and thus the above system has a solution and the polynomial
  	$Q$ exists.
  	Furthermore, we can compute $\bd=A^{-1}\bg=\frac{1}{\det(A)}C\cdot \bg$,
	where $C$ is the cofactor matrix of $A$.
	Since all entries of $A$ are bounded by $O(1)$,
	then the entries of $C$, being cofactors of $A$,
	are also bounded by $O(1)$.
  	Since $|\bg_k|=O(w)$ and $|I|=\omega(\frac{1}{\eta\tau})$,
  	for every $k=1,2,\cdots,\bb$,
	we have $|\bd_{ij}|=O(w/|I|^{\bbb})=o(w(\eta\tau)^{\bbb})=o(\delta\tau^{\bbb})$.
	
	However, since both $Z(P_1)$ and $Z(Q)$ pass through these $\bb$ points,
	both $P_1$ and $Q$ should satisfy $A\cdot \bc_1=0$ and $A\cdot \bc_2=0$,
	where $\bc_1, \bc_2$ are their coefficient vectors respectively.
	But since $\det(A)\neq 0$, $\bc_1=\bc_2$,
	meaning, $P_1\equiv Q$.
	This means for every $i,j=0,1,\cdots,\Delta$,
	where $j\neq\D$ and $i+j\le\Delta$,
	$|a_{ij}-b_{ij}|=\bd_{ij}=o(\delta\tau^{\bbb})$.
	However, by assumption,
	if two polynomials are not equal,
	then there exists at least one $c_{ij}$
	such that they differ by at least $\delta\tau^{\bbb}$,
	a contradiction.
	So $|I|=O(\frac{1}{\eta\tau})$.
\end{proof}

We now generalize Lemma~\ref{lem:intlen} to higher dimensions.

\begin{lemma}
\label{lem:intmeasure}
	Let $P_1, P_2$ be two distinct distant $D$-variate polynomials. 
	Let $S=\{X:\pi(Z(P_1),Z(P_2),X)=O(w)\land X\in[1,2]^{D-1}\}$, where $w=\delta/\eta^{\bbb}=o(1)$.
	Then $\mu^{D-1}(S)=O(\frac{1}{\eta\tau})$.
\end{lemma}
\begin{proof}
	We prove the lemma by induction.
	The base case when $D=2$ is Lemma~\ref{lem:intlen}.
	Now suppose the lemma holds for dimension $D-1$, we prove it for dimension $D$.
	Observe that we can rewrite a $D$-variate polynomial 
	$
		P(X)=X_1-X_2^{\Delta}+\sum_{i\in I^D}A_iX^i
	$
	as
	$
		P(X)=X_1-X_2^{\Delta}+\sum_{j\in I_{:D-1}^D}(f_j(X_D))X_{:D-1}^j,
	$
	where
	$
		f_j(X_D) = \sum_{k=0}^{\D-\sigma(j)}A_{j\oplus k}X_D^k.
	$
	Consider two distinct distant $D$-variate polynomials $P(X)=X_1-X_2^{\Delta}+\sum_{i\in I^D}A_iX^i$
	and $Q(X)=X_1-X_2^{\Delta}+\sum_{i\in I^D}B_iX^i$.
	Let $f_j, g_j$ be the corresponding coefficients for $X_{:D-1}^j$.
	Note that there exists some $j$ such that $f_j\not\equiv g_j$ because $P_1, P_2$ are distinct. 
	Let $h_j(X_D)=f_j(X_D)-g_j(X_D)$ and observe that $h_j$ is a univariate polynomial in $X_D$.
	We show that the interval length of $X_D$ in which 
	$|h_j(X_D)| < \xi_{D-1}$ is upper bounded by $O(\frac{1}{\eta\tau})$ for any $h_j(X_D)\not\equiv 0$.
	Pick any $h_j(X_D)\not\equiv 0$ and note that 
	this means there exists at least one coefficient of $h_j(X_D)$ that is nonzero.
	By assumption,
	each coefficient of $h_j(X_D)$ has absolute value at least $\xi_D$ if not zero.
	If the constant term is the only nonzero term, then the interval length of $X_D$
	in which $|h_j(X_D)|<\xi_{D-1}$ is 0, 
	since $|h_j(X_D)|\ge \xi_D > \xi_{D-1}$ by definition.
	Otherwise by Lemma~\ref{lem:polylem},
	the interval length $|r|$ for $X_D$ in which $|h_j(X_D)| < \xi_{D-1}$
	is upper bounded by
	\[
		|r|=O\left(\left(\frac{\xi_{D-1}}{\xi_D}\right)^{1/\Delta}\right)
		=O\left(\left(\frac{1}{(\eta\tau)^{\Delta}}\right)^{1/\Delta}\right)
		=O\left(\frac{1}{\eta\tau}\right).
	\]
	Since the total number of different $j$'s is $\Theta(1)$, 
	the total number of $h_j(X_D)$ is then $\Theta(1)$.
	So the total interval length for $X_D$
	within which there is some nonzero $h_j(X_D)$ with $|h_j(X_D)|<\delta\tau_{D-1}$
	is upper bounded by $\Theta(1)\cdot O(\frac{1}{\eta\tau}) = O(\frac{1}{\eta\tau})$.
	Since we are in a unit hypercube, we can simply upper bound $\mu^{D-1}(S)$
	by $O(\frac{1}{\eta\tau})\cdot\Theta(1)=O(\frac{1}{\eta\tau})$.
	Otherwise, by the inductive hypothesis, the $(D-2)$-measure of $S$ in $[1,2]^{D-2}$ 
	is upper bounded by $O(\frac{1}{\eta\tau})$.
	Integrating over all $X_D$, $\mu^{D-1}(S)$ is bounded by $O(\frac{1}{\eta\tau})$
	in this case as well.
\end{proof}

\subsection{Lower Bound for General Polynomial Slabs}
Now we are ready to present our lower bound construction.
We will use a set $\sS$ of $D$-variate polynomials in $\bR[X]$ of form:
\[
	P(X)=X_1-X_2^{\Delta}+\sum_{i\in I^D}A_iX^i,
\]
where $X$ is a $D$-tuple of indeterminates,
$I^D$ is an index set
containing all $D$-tuples $i$ satisfying $\sigma(i)\le\Delta$,
and each $A_i \in \{ k\xi_D:k=\lfloor\frac{\epsilon}{2\xi_D}\rfloor, \lfloor\frac{\epsilon}{2\xi_D}\rfloor+1,\cdots,\lfloor\frac{\epsilon}{\xi_D}\rfloor\}$
for some $\xi_D=\delta\tau^{\bbb}(\eta\tau)^{(D-2)\D}$ to be set later,
except for one special coefficient:
we set $A_{i}=0$
for $i=(0,\D,0,\cdots,0)$.
Note that every pair of the polynomials in $\sS$ is distant.
A general polynomial slab is defined to be a triple $(P,0,w)$ where $P\in\sS$
and $w$ is a parameter to be set later. 
We need $w=o(\epsilon)$
and $\epsilon=o(1)$.
% we set k to be in this range such that P-w is a packed polynomial...

We consider a unit cube $\U^D=\prod_{i=1}^D[1,2]\subseteq\bR^D$ and use Framework~\ref{thm:rrfw}.
Recall that to use Framework~\ref{thm:rrfw}, 
we need to lower bound the intersection $D$-measure of each slab we generated and $\U^D$,
and upper bound the intersection $D$-measure of two slabs.

Given a slab $(P,0,w)$ in our construction,
first note that both $P$ and $P-w$ are packed polynomials.
We define the width of $(P,0,w)$ to be the distance between $Z(P)$ and $Z(P-w)$ along the $X_1$-axis.
The following lemma shows that the width of each slab we generate will be $\Theta(w)$ in $\U^D$.
See Appendix~\ref{sec:proof-slabwidth} for the proof.

\begin{restatable}{lemma}{slabwidth}\label{lem:slabwidth}
	Let $P_1 \in\sS$ and $P_2=P_1-r$ for any $0\le r = O(w)$.
	Then $\pi(Z(P_1),Z(P_2),X)=\Theta(r)$ for any $X\in[1,2]^{D-1}$.
\end{restatable}

The following simple lemma bounds the $(D-1)$-measure of the projection of 
the intersection of the zero set of any polynomial in our construction and $\U^D$
on the $(D-1)$-dimensional subspace perpendicular to $X_1$-axis.
See Appendix~\ref{sec:proof-base} for the proof.

\begin{restatable}{lemma}{base}\label{lem:base}
	Let $P\in\sS$. The projection of $Z(P)\cap \U^D$ on the $(D-1)$-dimensional
	space perpendicular to the $X_1$-axis has $(D-1)$-measure $\Theta(1)$.
\end{restatable}

Combining Lemma~\ref{lem:slabwidth} and Lemma~\ref{lem:base},
we easily bound the intersection $D$-measure of any slab in our construction and $\U^D$.
\begin{corollary}
\label{cor:genslabarea}
	Any slab in our construction intersects $\U^D$ with $D$-measure $\Theta(w)$.
\end{corollary}

Combining Lemma~\ref{lem:slabwidth} and Lemma~\ref{lem:intmeasure},
we easily bound the intersection $D$-measure of two slabs in our construction in $\U^D$.
\begin{corollary}
\label{cor:genslabintarea}
	Any two slabs in our construction intersect with $D$-measure $O(\frac{w}{\eta\tau})$ in $\U^D$.
\end{corollary}

Since there are at most $\bm=\binom{D+\Delta}{D}-1$ parameters
for a degree-$\Delta$ $D$-variate monic polynomial,
the number of polynomial slabs
we generated is then
\[
	\Theta\left(\left(\frac{\epsilon}{\xi_D}\right)^{\bm}\right)
	=\Theta\left(\left(\frac{n}{Q(n)^{1+2\bbb+(D-2)\Delta}2^{((D-2)\Delta+2\bbb)\sqrt{\log n}}}\right)^{\bm}\right)
	=O(n^{\bm}),
\]
by setting $\delta=wQ(n)^{\bbb}$, $\eta=Q(n)$, $\tau=2^{\sqrt{log n}}$, $\epsilon=\frac{1}{Q(n)^{\bbb}2^{\bbb\sqrt{\log n}}}$, and
$w=c_wQ(n)/n$ for a sufficiently large constant $c_w$.
We pick $c_w$ s.t. each slab intersects $\U^D$ with $D$-measure, by Corollary~\ref{cor:genslabarea}, $\Omega(w)\ge 4\bm Q(n)/n$.
By Corollary~\ref{cor:genslabintarea} the $D$-measure of the intersection of two slabs is upper bounded by 
$O(\frac{w}{Q(n)2^{\sqrt{\log n}}}) = O(\frac{1}{n2^{\sqrt{\log n}}})$.
By Theorem~\ref{thm:rrfw}, we get the lower bound
$
	S(n) = \domega\left( n^{\bm} / Q(n)^{\bm+2\bm\bbb+\bm(D-2)\Delta-1} \right).
$
Thus we get the following result.

\begin{theorem}
\label{thm:genpolyslablb}
	Let $\sP$ be a set of $n$ points in $\bR^D$, where $D\ge 2$ is an integer.
	Let $\sR$ be the set of all $D$-dimensional generalized polynomial slabs $\{(P, 0, w):\deg(P)=\Delta \ge 2, w > 0\}$
	where $P\in\bR[X_1,X_2,\cdots, X_D]$ is a monic degree-$\Delta$ polynomial.
	Let $\bb$ (resp. $\bm$) be the maximum number of parameters needed to specify a moinc degree-$\D$
	bivariate (resp. $D$-variate) polynomial.
	Then any data structure for $\sP$ that can answer generalized polynomial slab reporting
	queries from $\sR$ with query time $Q(n) + O(k)$,
	where $k$ is the output size, must use 
	$S(n)=\domega\left( \frac{n^{\bm}}{Q(n)^{\bm+2\bm\bbb+\bm(D-2)\Delta-1}}\right)$ space, where
	and $\bbb=\binom{\bb}{2}$.
\end{theorem}
\section{Data Structures for Uniform Random Point Sets}

In this section, we present data structures
for an input point set $\sP$ uniformly randomly distributed in a unit square $\U=[0,1]\times[0,1]$ 
for semialgebraic range reporting queries in $\bR^2$.
Our hope is that some of these ideas can be generalized to build more efficient data
structures for general point sets. 
To this end, we show two approaches based on two different assumptions:
one assumes the query curve has bounded curvature,
and the other assumes bounded derivatives.
We show that for any degree-$\D$ bivariate polynomial inequality,
we can build a data structure with space-time tradeoff $S(n)=\tilde O(n^{\bm}/Q(n)^{3\bm-4})$,
which is optimal for $\bm=3$~\cite{afshani2021lower}.
When the query curve has bounded derivatives for the first $\D$ orders within $\U$,
this bound sharpens to $\tilde O(n^{\bm}/Q(n)^{((2\bm-\D)(\D+1)-2)/2})$,
which matches the lower bound in~\cite{afshani2021lower} for polynomial slabs
generated by inequalities of form $y-\sum_{i\le\D}a_ix^i\ge0$.
Since any polynomial can be factorized into a product of $O(1)$ irreducible polynomials,
and we can show that any irreducible polynomial has bounded curvature (See Appendix~\ref{sec:irreducible} for details),
we can express the original range by a semialgebraic set consisting of $O(1)$ irreducible polynomials.
We mention that both data structures can be made multilevel,
then by the standard result of multilevel data structures, see e.g.,~\cite{matousek1993range} or~\cite{agarwal2017simplex},
it suffices for us to focus on one irreducible polynomial inequality.
So the curvature-based approach works for all semialgebraic sets.
For both approaches, the main ideas are similar:
we first partition $\U$ into a $Q(n)\times Q(n)$ grid $G$,
and then build a set of slabs in each cell of $G$ to cover the boundary $\partial\rR$ of a query range $\rR$.
The boundaries of each slab consist of the zero sets of lower degree polynomials.
We build a data structure to answer degree-$\D$ polynomial inequality queries inside each slab,
then use the boundaries of slabs to express the remaining parts of $\rR$.
This lowers the degree of query polynomials,
and then we can use fast-query data structures to handle the remaining parts.
We assume our data structure can perform common algebraic operations in $O(1)$ time,
e.g., compute roots, compute derivatives, etc.

\subsection{A Curvature-based Approach}
\begin{figure}[h]
	\centering
	\includegraphics[width=0.3\textwidth]{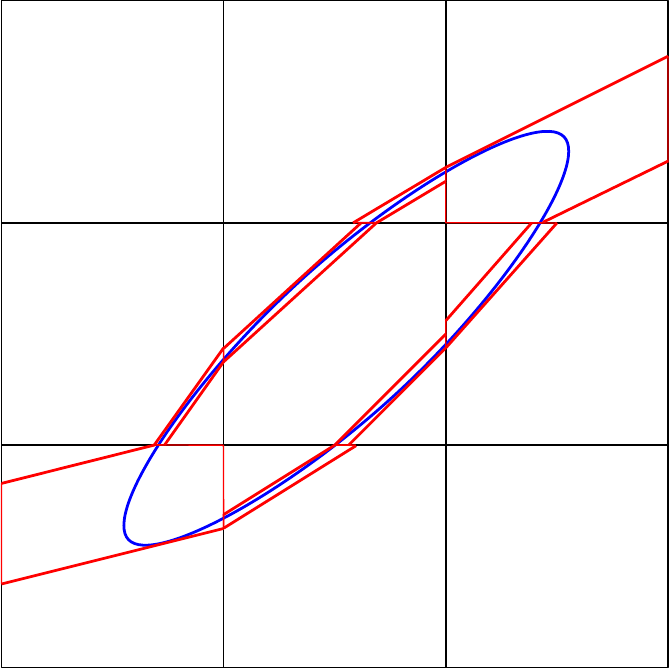}
	\caption{Cover an Ellipse with Slabs of Different Widths}
	\label{fig:ellipsecover}
\end{figure}

The main observation we use is that when the total absolute curvature of $\partial\rR$ is small,
the curve behaves like a line, and so we can cover it using mostly ``thin'' slabs,
and a few ``thick'' slabs when the curvature is big.
See Figure~\ref{fig:ellipsecover} for an example.
We use the curvature as a ``budget'':
thin slabs have few points in them so we can afford to store them in a ``fast'' data structure and 
the overhead will be small. 
Doing the same with the thick slabs will blow up the space too much so instead we store them in ``slower'' 
but ``smaller'' data structures. 
The crucial observation here is that for any given query, we only need to use a few ``thick'' slabs
so the slower query time will be absorbed in the overall query time.

The high-level idea is to build a two-level data structure.
For the bottom-level, we build a multilevel simplex range reporting data structure~\cite{matousek1993range}
with query time $\tilde O(1) + O(k)$ and space $S(n)=\tilde O(n^2)$.
For the upper-level,
for each cell $C$ in $G$ and a parameter $\alpha = 2^i/Q(n)$, for $i=0, \cdots, \lfloor\log Q(n) \rfloor$,
we generate a series of parallel disjoint slabs of width $\alpha/Q(n)$ such
that they together cover $C$.
Then we rotate these slabs by angle $\gamma = j/Q(n)$, for $j=1,2,\cdots,\lfloor 2\pi  Q(n) \rfloor$.
For each slab we generated during this process,
we collect all the points in it and build a $\tilde O(Q(n)\alpha)+O(k)$ query time
and $\tilde O((n/(Q(n)\alpha))^{\bm})$ space data structure
by linearization~\cite{yao1985ageneral} to $\bR^{\bm}$ and using simplex range reporting~\cite{matousek1993range}.

The following lemma shows we can efficiently report the points close to $\partial\rR$ using slabs we constructed.
For the proof of this lemma, we refer the readers to Appendix~\ref{sec:proof-cellslabcover}.
\begin{restatable}{lemma}{cellslabcover}\label{lem:cellslabcover}
	We can cut $\partial\rR$ into a set $\sS$ of $O(Q(n))$ sub-curves 
	such that for each sub-curve $\sigma$,
  we can find a set $S_{\sigma}$ of slabs that together cover $\sigma$.
  Let $P_{\sigma}$ be the subset of the input that lies inside the query
  and inside the slabs, i.e., 
  $P_{\sigma} = \rR \cap\sP \cap(\cup_{s\in S_{\sigma}}s)$.
  $P_{\sigma}$  can be reported in time 
  $Q(n)\tilde O(\kappa_{\sigma}+1/Q(n)) + O(|P_{\sigma}|)$,
  where $\kappa_{\sigma}$ is the total absolute curvature of $\sigma$.
	Furthermore, for any two distinct $\sigma_1,\sigma_2\in\sS$,
	$s_1\cap s_2=\emptyset$ for all $s_{1}\in S_{\sigma_1}, s_{2}\in S_{\sigma_2}$.

\end{restatable}

With Lemma~\ref{lem:cellslabcover}, we can bound the total query time for points close to $\partial\rR$ by
$
	\sum_{\sigma} Q(n) \tilde O(\kappa_{\sigma}+1/Q(n)) + O(t_{\sigma}) = \tilde O(Q(n)) + O(t_1),
$
where $t_1$ is the output size. 
An important observation is that after covering $\partial\rR$,
we can express the remaining regions by the boundaries of the slabs used and $G$,
which are linear inequalities and so we can use simplex range reporting.
Lemma~\ref{lem:linregion} characterizes the remaining regions. See Appendix~\ref{sec:proof-linregion} for the proof.
\begin{restatable}{lemma}{linregion}\label{lem:linregion}
	There are $O(Q(n))$ remaining regions and each region can be expressed using $O(1)$ linear inequalities. 
	These regions can be found in time $O(Q(n))$.
\end{restatable}
With Lemma~\ref{lem:linregion}, the query time for the remaining regions is $\tilde O(Q(n)) + O(t_2)$,
where $t_2$ is the number of points in the remaining regions.
Then the total query time is easily computed to be bounded by $\tilde O(Q(n)) + O(k)$, where $k=t_1+t_2$.

To bound the space usage for the top-level data structure,
note that we have $Q(n)^2$ cells, for each $\alpha$, we generate $\Theta(\frac{1/Q(n)}{\alpha/Q(n)})=\Theta(1/\alpha)$ slabs
for each of the $\Theta(Q(n))$ angles.
Since points are distributed uniformly at random,
the expected number of points in a slab of width $\alpha/Q(n)$ in a cell $C$
is $O(n\cdot\frac{1}{Q(n)}\cdot\frac{\alpha}{Q(n)})$.
So the space usage for the top-level data structure is
\[
	S(n) = \sum_{\alpha} Q(n)^2 \cdot \Theta\left(\frac{1}{\alpha}\right) \cdot \Theta(Q(n)) \cdot \tilde O\left(
			\frac{O\left(
				n\cdot\frac{1}{Q(n)}\cdot\frac{\alpha}{Q(n)}
				\right)
			}{
				Q(n)\alpha
			}
		\right)^{\bm}
		= \tilde O
		\left(
			\frac{n^{\bm}}{Q(n)^{3\bm-4}}
		\right).
\]
On the other hand, we know that the space usage for the bottom-level data structure is $\tilde O(n^2)$.
%For technical reasons, we need an $O(Q(n)^4)$ structure to help find slabs. See Lemma~\ref{lem:gadget} in Appendix~\ref{sec:proof-cellslabcover}.
So the total space usage is bounded by $\tilde O(\frac{n^{\bm}}{Q(n)^{3\bm-4}})$ for $\bm\ge3$.

We therefore obtain the following theorem.
\begin{theorem}
\label{thm:semigends}
	Let $\sR$ be the set of semialgebraic ranges formed by degree-$\D$ bivariate polynomials.
	Suppose we have a polynomial factorization black box
	that can factorize polynomials into the product of irreducible polynomials in time $O(1)$,
	then for any $\log^{O(1)}n \le Q(n) \le n^{\epsilon}$ for some constant $\epsilon$, 
	and a set $\sP$ of $n$ points distributed uniformly randomly in $\U=[0,1]\times[0,1]$,
	we can build a data structure of space $\tilde O(n^{\bm}/Q(n)^{3\bm-4})$
	such that for any $\rR\in\sR$, we can report $\rR\cap\sP$ in time $\tilde O(Q(n)) + O(k)$ in expectation,
	where $\bm \ge 3$ is the number of parameters 
	needed to define a degree-$\D$ bivariate polynomial and $k$ is the output size.
\end{theorem}

\subsection{A Derivative-based Approach}

If we assume that the derivative of $\partial\rR$ is $O(1)$,
the previous curvature-based approach can be easily adapted to get a derivative-based data structure.
See Appendix~\ref{sec:simpleds} for details.
We can even do better by using slabs whose boundaries are the zero set of higher degree polynomials
instead of linear polynomials.
Using Taylor's theorem, we show that we can cover the boundary of the query using ``thin'' slabs of lower degree
polynomials, similar to the approach above.
The full details are presented in Appendix~\ref{sec:multiderds}.

\begin{theorem}
\label{thm:hdds}
	Let $\sR$ be the set of semialgebraic ranges formed by degree-$\D$ bivariate polynomials
	with bounded derivatives up to the $\D$-th order.
	For any $\log^{O(1)}n \le Q(n) \le n^{\epsilon}$ for some constant $\epsilon$,
	and a set $\sP$ of $n$ points distributed uniformly randomly in $\U=[0,1]\times[0,1]$,
	we can build a data structure which uses space $\tilde O(n^{\bm}/Q(n)^{((2\bm-\D)(\D+1)-2)/2})$
	s.t. for any $\rR\in\sR$, we can report $\sP\cap\rR$ in time $\tilde O(Q(n)) + O(k)$ in expectation,
	where $\bm$ is the number of parameters needed to define a degree-$\D$ bivariate polynomial
	 and $k$ is the output size.
\end{theorem}

\begin{remark}
	We remark that our data structure can also be adapted to support semialgebraic range searching queries in the semigroup model.
\end{remark}

\section{Conclusion and Open Problems}

In this paper,
we essentially closed the gap between the lower and upper bounds
of general semialgebraic range reporting in the fast-query case
at least as far as the exponent of $n$ is concerned.
We show that for general polynomial slab queries
defined by $D$-variate polynomials of degree at most $\Delta$ in $\bR^D$
any data structure with query time $n^{o(1)}+O(k)$
must use at least $S(n)=\domega(n^{\bm})$ space,
where $\bm=\binom{D+\D}{D}-1$ is the maximum possible
parameters needed to define a query.
This matches current upper bound (up to an $n^{o(1)}$ factor).

We also studied the space-time tradeoff
and showed an upper bound that
matches the lower bounds in~\cite{afshani2021lower}
for uniform random point sets.

The remaining big open problem here is proving a tight bound for the exponent of $Q(n)$ in the space-time tradeoff.
There is a large gap between the exponents in our lower bound  versus
the general upper bound.
Our results show that current upper bound might not be tight.
On the other hand,
our lower bound seems to be suboptimal
when the query time is $n^{\Omega(1)}+O(k)$.
Both problems seem quite challenging,
and probably require new tools.

\bibliography{reference}

%\section{Appendices}

\appendix

\section{Proof of Theorem~\ref{thm:rrfw}}
\label{sec:proof-rrfw}
\rrfw*

First we present the original lower bound framework by
Chazelle~\cite{chazelle1990loweri} and Chazelle and Rosenberg~\cite{chazelle1996simplex}.

\begin{theorem}
\label{thm:chazellefw}
	Suppose the $D$-dimensional geometric range reporting problems
	admit an $S(n)$ space $Q(n)+O(k)$ query time data structure,
	where $n$ is the input size and $k$ is the output size.
	Assume we can find $m$ subsets $q_1, q_2, \cdots, q_m \subset \sS$
	for some input point set $\sS$,
	where each $q_i, i=1, \cdots, m$ is the output of some query
	and they satisfy the following two conditions:
	(i) for all $i=1,\cdots,m$, $|q_i|\ge Q(n)$; and
	(ii) $|q_{i_1} \cap q_{i_2} \cap \cdots q_{i_{\alpha}}| \le c$
	for some value $c \ge 2$.
	Then, we have $S(n)=\Omega(\frac{\sum_{i+1}^m|q_i|}{\alpha 2^{O(c)}})=\Omega(\frac{mQ(n)}{\alpha 2^{O(c)}})$.
\end{theorem}

A common way to use this framework is through a ``volume'' argument,
i.e., we generate a set of geometric ranges in a hypercube and then show that
they satisfy the following two properties:
\begin{itemize}
\item Each range intersects the hypercube with large Lebesgue measure;
\item The Lebesgue measure of the intersection of any $k$ ranges is small.
\end{itemize}
Then if we sample $n$ points uniformly at random in the hypercube,
we obtain $\sS$ in~\autoref{thm:chazellefw} in expectation.
However, we generally want to show a lower bound for the worst case,
then we need a way to derandomize to turn the result to a worst-case lower bound.
We now introduce some derandomization techniques,
which are direct generalizations of the $2$D version of the derandomization lemmas in~\cite{afshani2021lower}.
Given a $D$-dimensional hypercube $\C^D$ of side length $|l|$ and a set of ranges.
The first lemma shows that when each range intersects $\C^D$ with large $D$-dimensional Lebesgue measure
(For simplicity, we will call such a measure $D$-measure and denoted by $\mu^D(\cdot)$.)
and the number of ranges is not too big,
then with high probability, each range will contain many points.

\begin{lemma}
\label{lem:derandslab}
	Let $\C^D$ be a hypercube of side length $|l|$ in $\bR^D$.
	Let $\sR$ be a set of ranges in $\C^D$ satisfying two following conditions:
	(i) the $D$-measure of the intersection of any range $\rR \in \sR$ and $\C^D$ is at least $c|l|^Dt/n$ for some constant $c \ge 4k$
	and a parameter $t \ge \log n$ for some value $k \ge 2$;
	(ii) the total number of ranges is bounded by $O(n^{k+1})$.
	Now if we sample a set $\sP$ of $n$ points uniformly at random in $\C^D$,
	then with probability $> 1/2$, $|\sP\cap\rR|\ge t$ for all $\rR\in\sR$.
\end{lemma}
\begin{proof}
	We pick $n$ points in $\C^D$ uniformly at random.
	Let $X_{ij}$ be the indicator random variable with
	\[
		X_{ij}=
	\begin{cases}
		1, \textrm{point $i$ is in range $j$},\\
		0, \textrm{otherwise.}
	\end{cases}
	\]
	Since $\mu^D(\rR) \ge c|l|^Dt/n$ for every $\rR\in\sR$,
	the expected number of points in each range is at least $ct$.
	Consider an arbitrary range, let $X_j=\sum_{i=1}^nX_{ij}$, then by Chernoff's bound
	\begin{align*}    
		\Pr&\left[X_j<\left(1-\frac{c-1}{c}\right)ct\right]<e^{-\frac{\left(\frac{c-1}{c}\right)^2ct}{2}}\\
		\implies\Pr&[X_j<t]<e^{-\frac{(c-1)^2t}{2c}}<\frac{1}{n^{\frac{(c-1)^2}{2c}}}\le\frac{1}{n^{2k-1+1/(8k)}},
	\end{align*}
	where the second last inequality follows from $t\ge \log n$ and the last inequality follows from $c\ge4k$.
	Since the total number of ranges $O(n^{k+1})$,
	by the union bound, for $k\ge 2$ and a sufficiently large $n$,
	with probability $>\frac{1}{2}$,
	$|\sP\cap\rR|\ge t$ for all $\rR\in\sR$.
\end{proof}

The second lemma tells a different story: when the $D$-measure of the intersection of any $k$ ranges is small,
and the number of intersection is not too big,
then with high probability, each intersection has very few points.

\begin{lemma}
\label{lem:derandint}
	Let $\C^D$ be a hypercube of side length $|l|$ in $\bR^D$.
	Let $\sR$ be a set of ranges in $\C^D$ satisfying the following two conditions:
	(i) the $D$-measure of the intersection of any $t\ge 2$ distinct ranges $\rR_1, \rR_2, \cdots, \rR_t \in \sR$
	is bounded by $O(|l|^{D}/(n2^{\sqrt{\log n}}))$;
	(ii) the total number of intersections is bounded by $O(n^{2k})$ for $k \ge 1$.
	Now if we sample a set $\sP$ of $n$ points uniformly at random in $\C^D$,
	then with probability $> 1/2$,
	$|\rR_1 \cap \rR_2 \cap \cdots \cap \rR_t \cap \sP|<3k\sqrt{\log n}$ for all distinct ranges $\rR_1, \rR_2, \cdots,\rR_t\in\sR$. 
\end{lemma}
\begin{proof}
	We consider the intersection $\rho\in\C^D$ of any $t$ ranges and let $A=\mu^D(\rho)$.
	Let $X$ be an indicator random variable with
	\[
		X_i=
		\begin{cases}
			1, \textrm{the $i$-th point is inside $\rho$},\\
			0, \textrm{otherwise.}
		\end{cases}
	\]
	Let $X=\sum_{i=1}^nX_i$.
	Clearly, $\mathbb{E}[X] = \frac{An}{|l|^D}$.
	By Chernoff's bound,
	\[
		\Pr\left[X\ge(1+\delta)\frac{An}{|l|^D}\right]<\left(\frac{e^\delta}{(1+\delta)^{1+\delta}}\right)^{\frac{An}{|l|^D}},
	\]
	for any $\delta>0$.
	Let $\tau=(1+\delta)\frac{An}{|l|^D}$, then
	\[
		\Pr[X\ge\tau]<\frac{e^{\delta\frac{An}{|l|^D}}}{(1+\delta)^\tau}<\frac{e^\tau}{(1+\delta)^\tau}=\left(\frac{eAn}{|l|^D\tau}\right)^\tau.
	\]
	Let $\tau =3k\sqrt{\log n}$, since $A\le c|l|^D/(n2^{\sqrt{\log n}})$ for some constant $c$, we have
	\[
		\Pr\left[X\ge3k\sqrt{\log n}\right]<\left(\frac{ce}{2^{\sqrt{\log n}}3k\sqrt{\log n}}\right)^{3k\sqrt{\log n}}<\frac{(ce)^{3k\sqrt{\log n}}}{n^{3k}}.
	\]
	Since the total number of intersections is bounded by $O(n^{2k})$, the number of cells in the arrangement is also bounded
	by $O(n^{2k})$ and thus 
	by the union bound, for sufficiently large $n$,
	with probability $>\frac{1}{2}$,
	the number of points in every intersection region is less than $3k\sqrt{\log n}$.
\end{proof}

We now prove Theorem~\ref{thm:rrfw}.
\begin{proof}
	We sample a set $\sP$ of $n$ points uniformly at random in $\C^D$.
	Since each range $\rR_i$ has $\mu^D(\rR_i)\ge 4c |l|^DQ(n)/n$,
	and the number of ranges is $m=n^{c}$,
	then by Lemma~\ref{lem:derandslab},
	with probability more than $1/2$, $|\sP\cap\rR_i|\ge Q(n)$ for all $i=1,2,\cdots,m$.
	Since the intersection of any two ranges is upper bounded by $O(|l|^{D}/(n2^{\sqrt{\log n}}))$
	and the total number of intersections is $O(m^2)=O(n^{2c})$,
	then by Lemma~\ref{lem:derandint},
	with probability more than $1/2$,
	$|\rR_i\cap\rR_j\cap\sP|=O(\sqrt{\log n})$ for distinct ranges $\rR_i, \rR_j$.
	By the union bound,
	there is a point set such that both conditions in Theorem~\ref{thm:chazellefw} are satisfied,
	then we obtain a lower bound of
	\[
		S(n)=\Omega
			\left(
				\frac{mQ(n)}{2\cdot 2^{O(\sqrt{\log n})}}
			\right) 
			= 
			\domega(mQ(n)).\qedhere
	\]
\end{proof}

\section{Proof of Observation~\ref{obs:func}}
\label{sec:proof-func}
\func*
\begin{proof}
	We only need to show that there exists only one solution to equation
	$0=a_1-a_2^{\D}+f(a_1)$ when $a_1 > 0$ and the solution has value $O(1)$,
	where $f(a_1)$ is a polynomial in $a_1$ with nonnegative coefficients.
	Since $1\le a_2\le 2$, it easily follows.
\end{proof}

\section{Proof of Lemma~\ref{lem:slabwidth}}
\label{sec:proof-slabwidth}
\slabwidth*

\begin{proof}
	Pick any point $p=(p_1,p_2,\cdots,p_D)\in Z(P_1)$, and $p'=(p_1',p_2,\cdots,p_D)\in Z(P_2)$
	such that $p_i\in[1,2]$ for all $i=2,3,\cdots,D$,
	and $p_1' = p_1 + \gamma$.
	Clearly, $0<\gamma<1$ because $0\le r = O(w) = o(1)$.
	By definition
	\[
		P_2(p')=p_1 + \gamma + p_2^D + \left(\sum_{i}A_i(p_1+\gamma)^{i_1}p_2^{i_2}p_3^{i_3}\cdots p_D^{i_D}\right) -r
		= P_1(p) +\Theta(\gamma) - r = 0.
	\]
	So $\gamma=\Theta(r)$, meaning, $\pi(Z(P_1), Z(P_2), p)=\Theta(r)$ for $X\in[1,2]^{D-1}$.
\end{proof}

\section{Proof of Lemma~\ref{lem:base}}
\label{sec:proof-base}
\base*

We first bound the length of the $y$-interval within which a packed bivariate can intersect $\U^2$.

\begin{lemma}
\label{lem:curvelen}
	Let $P$ be a packed bivariate polynomial.
	Then $\sigma = Z(P)$ is fully contained in $\U^2$
	for some $y$-interval of length $\Theta(1)$.
\end{lemma}

\begin{proof}
	We show that $\sigma$ is sandwiched by curves
	$\sigma_l: x-y^{\Delta}+c\epsilon=0$ for some sufficiently large constant $c$
	and $\sigma_r: x-y^{\Delta}=0$ in $\U^2$.
	We intersect $\sigma_l, \sigma, \sigma_r$ with line $y=y_*$ for $y_*\in[1,2]$
	and denote the intersections to be $(x_l, y_*), (x_m, y_*), (x_r, y_*)$ respectively.
	Since $\sigma$ is of form $x-y^{\Delta}+\sum_{i=0}^{\Delta}\sum_{j=0}^{\Delta-i}c_{ij}x^iy^j=0$,
	$x_m=y_*^{\Delta}-O(\epsilon)$
	because $0\le c_{ij} = O(\epsilon)$ and $0<x=O(1)$ when $y_*\in[1,2]$ by Observation~\ref{obs:func}.
	So for sufficiently large $c$, $x_l\le x_m \le x_r$.
	It is elementary to compute that $\sigma_l$ and $\sigma_r$ intersect $x=1$
	at point $(1,\sqrt[\Delta]{1+c\epsilon}), (1,1)$ respectively, 
	and intersect $x=2$ at point $(2,\sqrt[\Delta]{2+c\epsilon}), (2,\sqrt[\Delta]{2})$ respectively in the first quadrant.
	So the intersection of $\sigma$ with $x=1$ (resp. $x=2$) has $y$-value between $1$ and $\sqrt[\Delta]{1+c\epsilon}$
	(resp. $\sqrt[\Delta]{2}$ and $\sqrt[\Delta]{2+c\epsilon}$).
	So the projection of $\sigma\cap\U^2$ onto the $y$-axis has length at least $\sqrt[\Delta]{2}-\sqrt[\Delta]{1+c\epsilon}$.
	Since $\epsilon=o(1)$, the lemma holds.
\end{proof}

Now we prove Lemma~\ref{lem:base}.
\begin{proof}
	We intersect $Z(P)$ with $X_i=a_i\in[1,2]$ for $i=3,4,\cdots,D$.
	The resulting polynomial will be a packed bivariate polynomial.
	By Lemma~\ref{lem:curvelen}, we know the intersection of the zero set
	of this bivariate polynomial and $\U^2$ has $1$-measure $\Theta(1)$ in the $X_2$-axis.
	Integrating over all $X_i$ for $i=3,4,\cdots,5$,
	$Z(P)$ intersects $\U^D$ with $(D-1)$-measure $\Theta(1)$
	in the subspace perpendicular to the $X_1$-axis.
\end{proof}

\section{Total Absolute Curvature of the Zero Set of Irreducible Polynomials}
\label{sec:irreducible}

In this section, we prove the following lemma.

\begin{lemma}
\label{lem:irreducible}
Let $P$ be an irreducible bivariate polynomial of constant degree.
	Then $Z(P)$ has total absolute curvature $O(1)$.
\end{lemma}

We first show for any value $v\in\bR\cup\{\pm\infty\}$,
the number of points on $Z(P)$ whose derivative achieves this value is $O(1)$.

We will use B\' ezout's Thoerem.
\begin{theorem}[B\' ezout's Theorem]\label{bezout}
  Given $2$ polynomials $P(x,y)$ and $Q(x,y)$ of degree $\Delta_p$ and $\Delta_q$
  respectively, either the number of common zeroes of $P$ and $Q$ is at most $\Delta_p\cdot \Delta_q$
  or they have a common factor. 
\end{theorem}

Now we show any irreducible polynomial has $O(1)$ points achieving the same derivative.

\begin{lemma}
\label{lem:polyder}
  Let $P(x,y)$ be an irreducible bivariate polynomial of degree $\Delta >1$.
	Then the number of points on $Z(P(x,y))$ which have a fixed derivative $c$ is bounded by $O(\Delta^2)$.
\end{lemma}
\begin{proof}
  For simplicity, we first rotate $P(x,y)$ such that the fixed derivative is $0$.
	Let us denote the new polynomial with $Q(x,y)$ and it is easy to see that $Q$ is also irreducible since
  if $Q$ could be written as $Q(x,y) = R(x,y)S(x,y)$, then $P(x,y)$ would also have a similar decomposition.
  
  By differentiating $Q$, we get that 
	$\frac{\dd y}{\dd x}= -\frac{Q_x(x,y)}{Q_y(x,y)}=0$, and thus 
	$Q_x(x,y)=0$.
  As a result, any point $(x,y)$ with derivative 0, lies on the zero set of $Q$ and $Q_x$. 

  Both $Q$ and $Q_x$ have degree $O(\Delta)$ and since $Q$ is irreducible  and degree of $Q_x$ is at least
  one, they cannot have a common factor. 
  By B\' ezout's Theorem, this implies that they have $O(\Delta^2)$ common zeroes. 
\end{proof}

We now prove Lemma~\ref{lem:irreducible}.
More specifically, we prove the following:
\begin{lemma}
\label{lem:anglebound}
  Consier a smooth curve $C$ such that 
	for any value $v$, there are at most $k$ points $p$ on $C$ such that the tangent line
  at $p$ has slope $v$.
  Then  $C$ has  total absolute curvature $O(k^2)$.
\end{lemma}
\begin{proof}	
	We parametrize $P(x,y)=0$ by its arc length $s$ over an interval $I$ and then consider 
	the function $\alpha:\bR\rightarrow \bR$ be a function that maps the arc length of the curve to the angle of the curve. 
  Note that $\alpha(s)$ is allowed to increase beyond $2\pi$.
  Let $\alpha_1$ and $\alpha_2$ be the infimum and surpremum of $\alpha(s)$ over $s \in I$.
  Note that we must have $\alpha_2 - \alpha_1\le k 2\pi$ as otherwise we can find more then $k$ points
  with the same slope on $C$.
  $\alpha'(s)$ determines the curvature of the curve at point $s$ and 
  its total curvature is 
  \[
    \int_I |\alpha'(s)| ds \le 2\pi k^2
  \]
  where the inequality follows from the observation that the equation 
  $\alpha(s) = v$ for every $v$ has at most $k$ solutions and thus the total
  change in $\alpha(s)$ is bounded by $k\cdot |\alpha_2 - \alpha_1| \le 2\pi k^2$.
\end{proof}

Lemma~\ref{lem:irreducible} then follows easily by Lemma~\ref{lem:polyder} and~\ref{lem:anglebound}.

\section{Proof of Lemma~\ref{lem:cellslabcover}}
\label{sec:proof-cellslabcover}
\cellslabcover*

Now suppose we have a sub-curve $\sigma\subset\partial\rR$ in $C$
that contains no singular points (points with undefined derivatives) except for possible the two boundaries,
if the total absolute curvature is between $0$ and $\pi/4$,
then we can efficiently find $O(1)$ slabs to cover it as shown in the following lemma.

\begin{lemma}
\label{lem:slabcover}
	Let $\sigma$ be any differentiable sub-curve in a cell $C$ with total absolute curvature $\kappa_\sigma$ such that $0\le\kappa_\sigma\le\pi/4$.
	We can find a set of $O(1)$ slabs of width $O(\kappa_{\sigma}/Q(n)+1/Q(n)^2)$ that together cover $\sigma$
	and these slabs can be found in time $\tilde O(1)$.
\end{lemma}

\begin{figure}[h]
  		\centering
  		\includegraphics[scale=0.5]{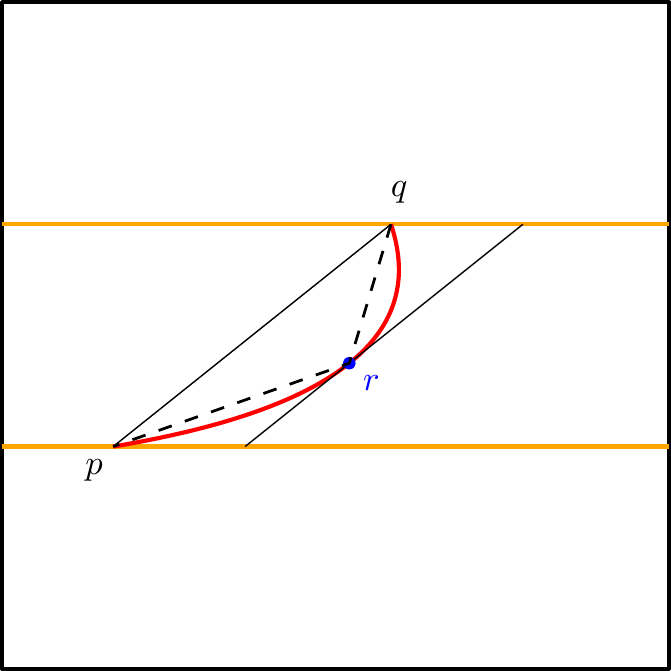}
  		\caption{Covering a Sub-curve Using Slabs}
  		\label{fig:curve-cover}
	\end{figure}

\begin{proof}
	Let $p$ and $q$ be the end points of the curve $\sigma$.
	Consider the point $r$ furthest away from the line $pq$ on the curve. 
	See Figure~\ref{fig:curve-cover} for an example.
  	Observe that we can use the mean value theorem 
	between $p$ and $r$ and also between $r$ and $q$.
  	This yields that the sum of the angles $\angle rpq + \angle rqp$ 
	is at most the total absolute curvature of $\sigma$. 
	Since $p,q$ are in $C$, 
	$|\overline{pq}|=O(1/Q(n))$ and since $\angle rpq, \angle rqp  \le \kappa_\sigma \le \pi/4$,
 	it follows that the distance between the line tangent to $r$ and $\overline{pq}$ is 
  	$O(\kappa_\sigma /Q(n))$.
  	Finally, notice that in our construction, we have created slabs of orientation $i/Q(n)$ for every integer $i$.
  	As a result,  we can cover $\sigma$ with $O(1)$ slabs of width $O(\kappa_{\sigma}/Q(n) + 1/Q(n)^2)$.
  	To find the slabs, we can use any of the previous techniques in semialgebraic range searching
  	since the input size (i.e., the number of slabs) in our construction is $Q(n)^{O(1)}$.
\end{proof}	

We now show how to decompose $\partial\rR$.
Observe that $\partial\rR$ intersects $O(Q(n))$ cells in $G$
because otherwise $\partial\rR$ will have $\omega(1)$ tangents,
which contradicts B\'ezout's theorem.
We cut $\partial\rR$ using these $O(Q(n))$ cells to get $\sS$.

Let $\sigma\subset\partial\rR$ be the sub-curve in a cell $C\in G$.
To find slabs to cover $\sigma$,
we  refine $\sigma$ to be smaller pieces of curves
to use Lemma~\ref{lem:slabcover}.
We simply cut $\sigma$ into pieces such that each piece has total absolute curvature $\le\pi/4$
and contains no singular points.
Recall that the singular points of the zero set of a bivariate polynomial
is a point where both partial derivatives are $0$.
By B\'ezout's theorem,
there are $O(1)$ singular points.
Since the total curvature of $\partial\rR$ is $O(1)$,
we will get $O(1)$ refined sub-curves.
This part is easy with the assumption of our model of computation
and so we omit the details about how to cut $\sigma$.

Now for each (refined) sub-curve $\sigma_r$, by Lemma~\ref{lem:slabcover} we can find $O(1)$ slabs to cover it.
We report points close to $\sigma_r$ as follows.
First we sort the slabs in some order.
Let $s$ be a slab we find for $\sigma_r$.
When we examine $s$, we use the data structure built in $s$
to find the points in $\rR$.
The query time will be $Q(n) \tilde O(\kappa_{\sigma_r}+1/Q(n)) + O(k)$
by Lemma~\ref{lem:slabcover}.
Before reporting the point,
we check if the point has been reported in slabs we have examined before.
This is because the slabs we found may intersect.
But since we have $O(1)$ refined sub-curves for $\sigma$ and each refined sub-curve requires $O(1)$ slabs to cover,
it takes only $O(1)$ time to check for duplicates.
Summing up the query cost for all refined sub-curves for $\sigma$,
the total query time is $Q(n) \tilde O(\kappa_{\sigma}+1/Q(n)) + O(t_{\sigma})$.
Since cells in $G$ are disjoint and each slab is built only for a specific cell,
the slabs we find for two distinct sub-curves will have zero intersection.
This proves Lemma~\ref{lem:cellslabcover}.

\section{Proof of Lemma~\ref{lem:linregion}}
\label{sec:proof-linregion}
\linregion*

There are two types of remaining regions.
First, cells fully contained in $\rR$ but do not intersect $\partial\rR$.
Second, the regions in a cell intersected by $\partial\rR$
but not covered by slabs.

\begin{figure}[h]
	\centering
	\includegraphics[width=0.8\textwidth]{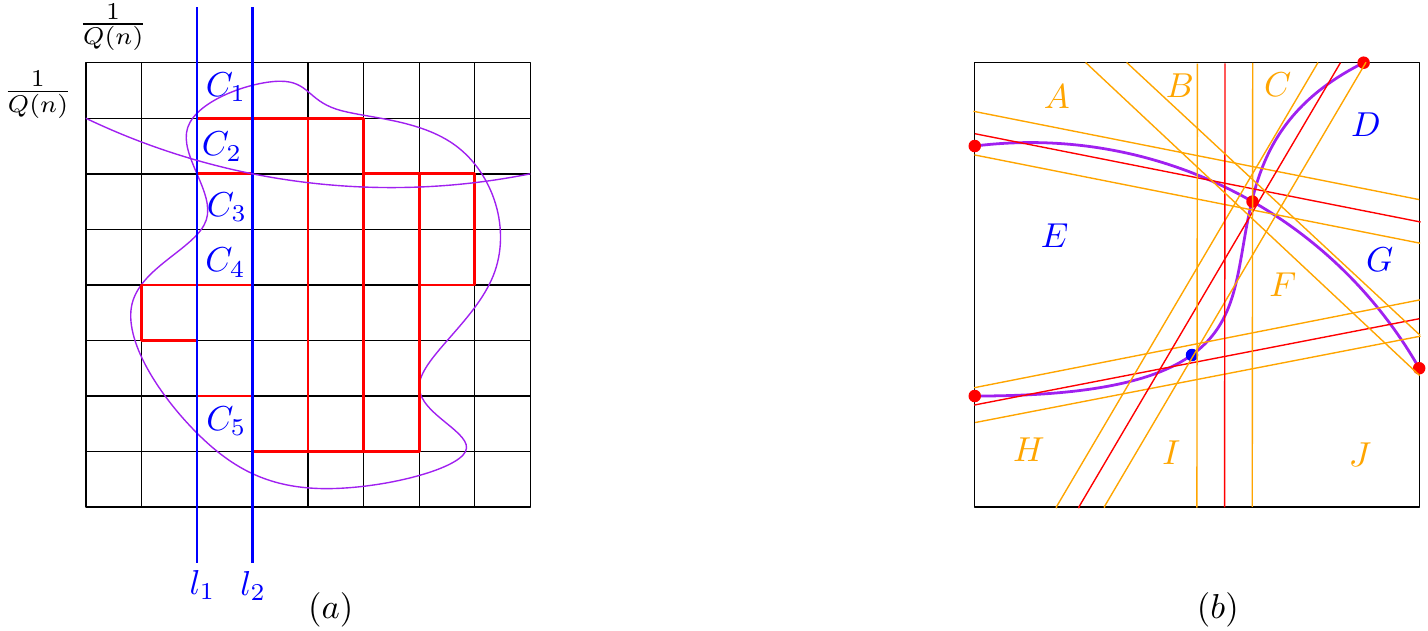}
	\caption{To Answer a Query: 
	$(a)$: Finding cells fully contained in $\partial\rR$.
	We have a chunk of zero cell between pairs $(C_1,C_2)$,
	$(C_2,C_3)$, and $(C_3,C_4)$,
	and a chunk of two cells between $C_4, C_5$.
	$(b)$: Covering a sub-curve $\sigma$ in a cell.
	Red dots are singular points of $\partial\rR$ 
	and its intersections with $C$.
	The blue dots is used to make sure each refined sub-curve has total absolute curvature $\le\pi/4$.
	We use slabs (denoted by orange/red line segments) 
	to cover the boundaries of $\sigma$.
	There are $10$ regions in the subdivision formed by the outmost boundaries of slabs.
	Three of them $(D,E,G)$ are fully contained in $\rR$.}
	\label{fig:simple-cover}
\end{figure}

We first handle the first type.
For any two adjacent vertical lines $l_1,l_2$ in the grid $G$,
we find all the cells between them
intersected by $\partial\rR$ in decreasing order
with respect to their $y$-coordinates.
For two consecutive cells $C_1, C_2$ we find,
all the cells between $C_1, C_2$
must be all contained or all not contained in $\rR$
because otherwise $C_1,C_2$ are not adjacent.
We then express the union of cells in between $C_1,C_2$
using four linear inequalities.
By this,
we can find all the cells intersecting $\partial\rR$
and all the chunks of cells fully contained in $\rR$
between $l_1,l_2$.
We do this for every consecutive pair of vertical lines.
The number of chunks is linear to the number of cells
intersecting $\partial\rR$ which is $O(Q(n))$ by B\'ezout's theorem,
so we have $O(Q(n))$ chunks as well.
See Figure~\ref{fig:simple-cover} (a) for an example.

For the second type,
observe that each such region is 
defined by the boundaries of $C$
(and/or) the outermost boundaries of slabs 
we used to cover sub-curves.
Since by the analysis of Lemma~\ref{lem:cellslabcover},
the sub-curve in a cell $C$ requires only $O(1)$ slabs to cover.
The outmost boundaries of these $O(1)$ slabs 
form a subdivision of complexity $O(1)$.
Since each face in the subdivision is either fully contained in $\rR$
or not contained in $\rR$,
it suffices to check an arbitrary point in the face.
We omit the details here.
In one cell, we have $O(1)$ remaining regions (faces in the subdivision)
and it takes $O(1)$ time to find it.
Since $\partial\rR$ intersects $O(Q(n))$ regions,
there are $O(Q(n))$ regions in total
and it takes $O(Q(n))$ time to find them.
See Figure~\ref{fig:simple-cover} (b) for an example.
This proves Lemma~\ref{lem:linregion}.

\section{An $S(n)=\tilde O(n^{\bm}-Q(n)^{3\bm-4})$ Derivative-based Data Structure}
\label{sec:simpleds}

The data structure is similar to the curvature-based one.
We also build a two-level data structure.
For each cell $C$,
we ``guess'' $Q(n)$ first derivatives $\alpha_1=-c, -c+t, -c+2t, \cdots, c$,
for $t=2c/Q(n)$.
For each guess $\alpha_1$,
we generate a series of disjoint parallel slabs of (vertical) width $w_v=1/Q(n)^2$ each
that together cover $C$
such that the boundary of each slab has derivative $\alpha_1$.
Since $|\alpha_1|=O(1)$,
the angle $\gamma$ between any slab and the $x$-axis is also $O(1)$,
so the width of each slab is $w=w_v\cdot\cos\gamma=\Theta(w_v)$.
Therefore the total number of slabs we generate for each $\alpha_1$ in a cell 
is $\frac{\Theta(1/Q(n))}{\Theta(w_v)}=O(Q(n))$.
%For each slab $s$,
%we call the intersection of the lower boundary of $s$
%and the induced line of the left boundary of $C$ the anchor of $s$.
For each $s$, we collect the points in it
and build an $\tilde O(1)+O(k)$ query time and $\tilde O(n^{\bm})$ space data structure.
This is our top-level data structure.
For the bottom-level data structure, we still use a multilevel simplex range reporting data structure
with $\tilde O(n^2)$ space and $\tilde O(1) + O(k)$ query time.

The space usage for the top level data structure is easily bounded to be
\[
	S_1(n) 
	= \tilde O
	\left( 
		Q(n)^2 \cdot \frac{2c}{2c/Q(n)} \cdot O(Q(n)) \cdot
		\left(
			\frac{1}{Q(n)^2}\cdot \frac{1}{Q(n)} \cdot n
		\right)^{\bm}
	\right)
	= \tilde O
	\left(
		\frac{n^{\bm}}{Q(n)^{3\bm-4}}
	\right).
\] 
Since the bottom level data structure takes up $\tilde O(n^2)$ space.
The total space usage is $\tilde O(\frac{n^{\bm}}{Q(n)^{3\bm-4}})$ for $\bm\ge3$.

For the query answering, we prove a lemma similar to Lemma~\ref{lem:slabcover}.

\begin{lemma}
\label{lem:simpleslabcover}
	In our construction,
	if some differentiable sub-curve $\sigma$
	is contained in $C$,
	then we can find $O(1)$ slabs
	that cover $\sigma$.
	The time needed to find all these slabs is $\tilde O(1)$.
\end{lemma}

\begin{proof}
	Let $(p_x, p_x), (q_x, q_y)$ be the left and right endpoints of $\sigma$
	and $\frac{\dd y}{\dd x} (p_x, p_y)=\alpha_1^*$.
	Let $f(x)$ be the implicit function defined by $\sigma$ in between $(p_x, p_y)$ and $(q_x,q_y)$.
	Let $g(x)=\alpha_1(x-p_x)+p_y$ be the line passing through $(p_x, p_y)$
	with slope $\alpha_1$.
	Define the vertical distance between $f(x)$ and $g(x)$ in $[p_x,q_x]$ to be
	$
		d(x)=f(x)-g(x).
	$
	Since we guess $\alpha_1 = \frac{\dd y}{\dd x}(p_x,p_y)$ with step size $2\pi/Q(n)$,
	\[
	\begin{aligned}
		d(x)
		& = f(x) - (\alpha_1 (x-p_x) + p_y)\\
		& \le f(x) - ((\alpha_1^* \pm 2c/Q(n)) (x-p_x) + p_y)\\
		& = (f(x)-\alpha_1^*(x-p_x) - p_y) \pm 2c/Q(n)(x-p_x)\\
		& = \frac{f^{(2)}(\xi)}{2!}(x-p_x)^2 \pm 2c/Q(n)(x-p_x),
	\end{aligned}
	\]
	for some constant $\xi$ between $p_x$ and $x$,
	where the last equality follows from Taylor's theorem.
	Since $x \in [p_x, q_x]$ and $|q_x-p_x|\le 1/Q(n)$ as they are in $C$
	and all the derivatives are bounded, $|d(x)|=O(1/Q(n)^2)$.
	Since each slab has vertical width $w_v=1/Q(n)^2$,
	we only need $O(1)$ slabs to cover $\sigma$.
	
	To find these slabs, 
	by a similar analysis as in Lemma~\ref{lem:slabcover},
	since there are only $Q(n)^{O(1)}$ slabs in total,
	we can build a simple $Q(n)^{O(1)}$ size searching data structure
	to find the $O(1)$ slabs in time $\tilde O(1)$.
\end{proof}

Having Lemma~\ref{lem:simpleslabcover} in hand,
the query process is essentially the same as the one for the curvature-based solution
and the analysis is also the same by replacing Lemma~\ref{lem:slabcover}
by Lemma~\ref{lem:simpleslabcover}.
We omit the deials and present the following theorem.

\begin{theorem}
\label{thm:simpleder}
	Let $\sR$ be the set of semialgebraic ranges formed by degree-$\D$ bivariate polynomials
	with bounded derivatives up to the $\Delta$-th order.
	For any $\log^{O(1)}n \le Q(n) \le n^{\epsilon}$ for some constant $\epsilon$, 
	and a set $\sP$ of $n$ points distributed uniformly randomly in $\U=[0,1]\times[0,1]$,
	we can build a data structure of space $\tilde O(n^{\bm}/Q(n)^{3\bm-4})$
	such that for any $\rR\in\sR$, we can report $\rR\cap\sP$ in time $\tilde O(Q(n)) + O(k)$ in expectation,
	where $\bm$ is the number of parameters 
	needed to define a degree-$\D$ bivariate polynomial and $k$ is the output size.
\end{theorem}

\begin{remark}
	Note that we actually only need bounded derivatives up to the second order in Theorem~\ref{thm:simpleder}.
\end{remark}

\section{An $S(n)=\tilde O(n^{\bm}/Q(n)^{((2\bm-\D)(\D+1)-2)/2})$ Derivative-based Data Structure}
\label{sec:multiderds}

Now we improve the results in Appendix~\ref{sec:simpleds}.
The main idea is to use slabs formed by 
higher degree polynomial equalities.
These slabs work as finer and finer approximations
to the boundaries of query ranges.
We first define some notations.
\begin{definition}
	Let $I_x=[x_l, x_r]$ be an interval in the $x$-axis.
	Let $U(x)$ and $L(x)$ be two degree-$i$ polynomials in $x$ such that $\forall x\in I_x,U(x)>L(x)$.
	We say that the region enclosed by $U(x)$, $L(x)$, $x=x_l$ and $x=x_r$ is an $i$-slab $s$.
	We also say the $x$-range of $s$ is $[x_l,x_r]$.
	Furthermore, if for all $x\in I_x$, $U(x)-L(x)=w$, we say $s$ is a uniform slab with width $w$.
\end{definition}

In our application, $L(x), U(x)$ will be two degree-$i$ polynomial functions
that differ only in their constant terms.
It is not hard to see that in this case,
all the slabs are in fact uniform.

In a nutshell, our data structure $\Psi_{\Delta}$ for degree-$\D$ polynomial inequalities is still a two-level data structure.
The top-level structure is similar to that we described in Appendix~\ref{sec:simpleds}
but instead of using $1$-slabs,
we use $(\Delta-1)$-slabs.
These $(\Delta-1)$-slabs will have width $1/Q(n)^{\Delta}$
and we build data structures of size $\tilde O(n^{\bm_{2,\D}})$ for
the points in each slab that can answer semialgebraic queries defined by
degree-$\Delta$ polynomial inequalities in $\tilde O(1) + O(k)$ time.
The second part is a data structure built for the entire input points and it can answer degree-$(\D-1)$
polynomial inequality queries in time $\tilde O(1) + O(k)$ with space usage $\tilde O(n^{\bm_{2,\D-1}})$.
The overall idea of our data structure is the following:
given $\rR$, we use $(\Delta-1)$-slabs to cover its boundary.
Then the remaining parts will be defined by degree-$(\Delta-1)$ polynomial inequalities.
So we can use the bottom-level data structure to solve them.

Now we describe the details.
We first describe how to generate $i$-slabs for $i=1,2,\cdots,\Delta-1$.
The base $1$-slabs are what we have described in Appendix~\ref{sec:simpleds}.
Now assume we already have an $(i-1)$-slab $s_{i-1}$,
we generate $i$-slabs as follows.
Let the $x$-range of $s_{i-1}$ be $[x_l,x_r]$.
Let $\alpha_j^l=\frac{\dd^j y}{\dd x^j}(x_l)$ for $j=1,2,\cdots,i-1$
be the $j$-th order derivatives of $L(x)$ of $s_{i-1}$ at $x=x_l$.
Now to construct $L(x)$ of an $i$-slab $s_i$,
we make $Q(n)$ finer guesses for each $\frac{\dd^j y}{\dd x^j}(x_l)$.
Specifically, $\frac{\dd^j y}{\dd x^j}(x_l)=\alpha_j^l, \alpha_j^l + \frac{2c}{Q(n)^{i-j+1}}, \alpha_j^l + 2\cdot\frac{2c}{Q(n)^{i-j+1}}, \cdots, \alpha_j^l + \frac{2c}{Q(n)^{i-j}}$,
for $j=1,2,\cdots,i-1$,
and $\frac{\dd^i y}{\dd x^i}(x_l) = -c + \frac{2c}{Q(n)},  -c + 2\cdot\frac{2c}{Q(n)}, \cdots,  c$.
We then place ``anchor'' points evenly spaced with distance $1/Q(n)^{i+1}$ on the left boundary of $s_{i-1}$.
Every two degree-$i$ polynomials passing through adjacent anchor points
having the same $\frac{\dd^j y}{\dd x^j}(x_l)$ for $j=1,2,\cdots, i$ defines an $i$ slab.
If any two degree-$i$ polynomials $P(x), Q(x)$ have the same
$k$-th derivatives for all $k=1,2,\cdots,i$
at two points $(x_l, y_1)$, $(x_l, y_2)$,
it is elementary to show that for all $x$,
$|P(x)-Q(x)|=|y_1-y_2|$.
So every $i$-slab is uniform and its width is $1/Q(n)^{i+1}$.  

To build $\Psi_{\Delta}$,
we first build $1$-slabs as we did in Appendix~\ref{sec:simpleds},
and then repeatedly applying the process
described in the previous paragraph to get degree-$(\D-1)$ slabs.
Then we build the $\tilde O(n^{\bm_{2,\D}})$ space data structure in each slab
as the top-level data structure,
and then build the $\tilde O(n^{\bm_{2,\D-1}})$ space data structure
for all input points as the bottom-level data structure.

Now we bound the space usage.
By the above procedure, for each $(i-1)$-slab, $i\ge 3$
we generate $Q(n)^{i-2}$ guesses for derivatives
for the first $i-2$ derivatives,
and $Q(n)$ guesses for the $(i-1)$-th derivative.
We have $\frac{1/Q(n)^{i-1}}{1/Q(n)^{i}}=Q(n)$
anchor points for the lower boundaries of slabs to pass through.
So in total, we generate $Q(n)^{i-2}\cdot Q(n)\cdot Q(n) = Q(n)^i$ many $(i-1)$-slabs
in an $(i-2)$-slab.
We know from Appendix~\ref{sec:simpleds}
that the number of $1$-slabs
is upper bounded by $O(Q(n)^4)$. 
Since we only build fast-query data structures in $(i-1)$-slabs,
the total space usage of all the structures built on $(i-1)$-slabs is then bounded by
\[
\begin{aligned}
	S_1(n) 
	&= O
	\left(
		Q(n)^4 \cdot 
		\left(
			\prod_{j=3}^i Q(n)^j \cdot
		\right) \cdot
		\left(
			\frac{1}{Q(n)^i} \cdot \frac{1}{Q(n)} \cdot n
		\right)^{\bm_{2,i}}
	\right)\\
	&= O
	\left(
		Q(n)^{(i+1)i/2+1}
		\cdot
		\frac{n^{\bm_{2,i}}}{Q(n)^{\bm_{2,i}(i+1)}}
	\right)\\
	&= O
	\left(
		\frac{n^{\bm_{2,i}}}{Q(n)^{((2\bm_{2,i}-i)(i+1)-2)/2}}
	\right).
\end{aligned}
\]
As mentioned before, the space usage of the bottom-level
data structure for $\Psi_i$ is $\tilde O(n^{\bm_{2,i-1}})$.
Then for query time $Q(n)=n^{\epsilon}$ where $\epsilon$ is some small constant,
the space usage of our entire data structure
$\Psi_i$ is bounded by $\tilde O(n^{\bm_{2,i}}/Q(n)^{((2\bm_{2,i}-i)(i+1)-2)/2})$.

For query answering, we first show the following lemma,
which is a generalization of Lemma~\ref{lem:simpleslabcover}.
The proof idea is similar to Lemma~\ref{lem:simpleslabcover},
the only difference is now we consider a Taylor polynomial of degree-$(\D-1)$
instead of $1$.

\begin{lemma}
\label{lem:hdslabcover}
	In our construction,
	if some differentiable sub-curve $\sigma$
	is contained in some cell $C$,
	then we can find up to $O(1)$ $(\Delta-1)$-slabs
	to cover $\sigma$.
	The time needed to find these slabs is $\tilde O(1)$.
\end{lemma}
\begin{proof}
	Let $(p_x, p_y), (q_x, q_y)$ be the left and right endpoints of $\sigma$
	and $\frac{\dd^i y}{\dd x^i}(p_x, p_y)=\alpha_i^*$ for $i=1,2,\cdots,\Delta-1$.
	Let $f(x)$ be the implicit function defined by $\sigma$ in $[p_x,q_x]$
	and let $g(x)$ be a degree-$\D$ polynomial whose first $\Delta$ derivatives
	agree with those of $f(x)$ at point $(p_x, p_y)$.
	By Taylor's theorem, 
	the vertical distance between $f(x)$ and $g(x)$ is easily calculated to be bounded
	by $O(1/Q(n)^{\Delta+1})$ in $[p_x,q_x]$.
	Next we bound the vertical distance between $g(x)$ and the best fitting polynomial in our construction.
	Let $(a,b)$ be the intersection of $g(x)$ with the line containing the left boundary of $C$.
	Let $h(x)=\sum_{i=1}^{\Delta-1}\frac{\alpha_i}{i!}(x-a)^i+b$ be 
	a degree-$(\Delta-1)$ polynomial passing through $(a, b)$
	with $i$-th order derivative being $\alpha_i$ at $x=a$.
	We define the vertical distance between $g(x)$ and $h(x)$ in this range to be
	$
		d(x)=g(x)-h(x).
	$
	
	Since we guess $\alpha_i = \frac{\dd^i y}{\dd x^i}$ at $x=a$ with step size $2c/Q(n)^{\Delta-i}$ in our construction,
	\begin{align*}
		d(x)
		& = g(x) - 
			\left(
				\sum_{i=1}^{\Delta-1} \frac{\alpha_i}{i!} (x-a)^i + b
			\right)\\
		& \le g(x) - 
			\left(
				\sum_{i=1}^{\Delta-1} 
					\left(
						\frac{\alpha_i^* \pm 2c/Q(n)^{\Delta-i}}{i!}
					\right) 
					(x-a)^i + b
			\right)\\
		& = 
			\left(
				g(x) - 
				\left(
					\sum_{i=1}^{\Delta-1} \frac{\alpha_i^*}{i!} (x-a)^i + b
				\right)
			\right) 
			\pm 
			\sum_{i=1}^{\Delta-1} \frac{2c/Q(n)^{\Delta-i}}{i!} (x-a)^i\\
		& = \frac{g^{(\Delta)}(\xi)}{\Delta!}(x-a)^{\Delta} \pm \sum_{i=1}^{\Delta-1} \frac{2c/Q(n)^{\Delta-i}}{i!} (x-a)^i
	\end{align*}
	for some constant $\xi$ between $a$ and $x$,
	where the last equality follows from Taylor's theorem.
	Since $x \in [a, q_x]$ and $|q_x-a|\le 1/Q(n)$
	and all the derivatives of $g(x)$ are bounded in $\U$, $|d(x)|=O(1/Q(n)^{\Delta})$.
	Then the distance between $f(x)$ and $h(x)$ is bounded by $O(1/Q(n)^{\Delta+1})+|d(x)|=O(1/Q(n)^{\D})$ in $[p_x,q_x]$.
	Since each $(\Delta-1)$-slab has width $1/Q(n)^{\Delta}$,
	so it takes $O(1)$ $(\Delta-1)$-slabs to cover $\sigma$.
	To find these slabs, 
	by a similar analysis as in Lemma~\ref{lem:slabcover},
	since there are only $Q(n)^{O(1)}$ slabs in total,
	we can build a simple $Q(n)^{O(1)}$ size searching data structure
	to find the $O(1)$ slabs in time $\tilde O(1)$.
\end{proof}

With Lemma~\ref{lem:hdslabcover} in hand,
the query algorithm is essentially the same as the data structure described in Appendix~\ref{sec:proof-cellslabcover}
except for one minor difference:
here when we answer query in some cell,
we find $(\Delta-1)$-slabs
and use the fast query data structure in it.
But now since the boundaries of slabs are degree-$(\Delta-1)$ polynomials,
we need to handle ranges defined by $(\Delta-1)$ polynomial inequalities instead of linear inequalities.
This can be handled by our bottom-level data structure.
By a similar analysis as in Appendix~\ref{sec:proof-cellslabcover},
we can find $O(Q(n))$ $(\D-1)$-slabs to cover $\partial\rR$.
We can then report all the points close to $\partial\rR$ in time $\tilde O(Q(n)) + O(k)$.
The remaining regions of $\rR$ are defined by $O(Q(n))$ boundaries of the slabs we used and $G$
by a similar analysis as in Appendix~\ref{sec:proof-linregion}.
We use the bottom-level data structure for this part
and again we need $\tilde O(Q(n)) + O(k)$ time to report the points.
In total, the query time is bounded by $\tilde O(Q(n)) + O(k)$.
This proves Theorem~\ref{thm:hdds}.

Specifically, for polynomial inequalities of form $y+\sum_{a_i}x^i\le0$
or $x+\sum_{a_i}y^i\le0$, where $a_i\in\bR$ and $0\le i\le \Delta$ is an integer,
we have:

\begin{theorem}
	For Semialgebraic range $\rR$ formed by polynomial inequalities of form
	$y+\sum_{a_i}x^i\le0$ or $x+\sum_{a_i}y^i\le0$, 
	where $a_i\in\bR$ and $0\le i\le \Delta$ is an integer,
	and any $\log^{O(1)}n \le Q(n) \le n^{\epsilon}$ for some constant $\epsilon$,
	if the $n$ input points are distributed uniformly randomly in a unit square $\U=[0,1]\times [0,1]$,
	we can build a data structure of space $\tilde O(n^{\Delta+1}/Q(n)^{(\D+3)\D/2})$
	that answers range reporting queries with $\rR$ in time $\tilde O(Q(n)) + O(k)$ in expectation,
	where $k$ is the number of points to report.
\end{theorem}

\end{document}